\documentclass[fleqn,usenatbib]{mnras}

\usepackage[T1]{fontenc}

\DeclareRobustCommand{\VAN}[3]{#2}
\let\VANthebibliography\thebibliography
\def\thebibliography{\DeclareRobustCommand{\VAN}[3]{##3}\VANthebibliography}

\def\thebibliography{\DeclareRobustCommand{\VAN}[3]{##3}\VANthebibliography}

\usepackage{graphicx,amsmath,amssymb,makecell}
\usepackage{newtxtext,newtxmath}

\title[Globular Cluster ID using Machine Learning]{Using Machine Learning to Identify Extragalactic Globular Cluster Candidates from Ground-Based Photometric Surveys of M87}

\author[E. Barbisan et al.]{
Emilia Barbisan,$^{1,2}$\thanks{E-mail: emilia.barbisan@mail.mcgill.ca (EB)}
Jeff Huang,$^{1,2}$
Kristen C. Dage$^{1,2}$
Daryl Haggard,$^{1,2}$
Robin Arnason,$^{3}$
\newauthor
Arash Bahramian,$^{4}$
William I. Clarkson,$^{5}$
Arunav Kundu,$^{6}$
Stephen E. Zepf$^{7}$
\\
$^{1}$Department of Physics, McGill University, 3600 University Street, Montr\'eal, QC H3A 2T8, Canada\\
$^{2}$McGill Space Institute, McGill University, 3550 University Street, Montr\'eal, QC H3A 2A7, Canada\\
$^{3}$Interface Fluidics, Ltd., 11421 Saskatchewan Dr NW, Edmonton, AB T6G 2M9, Canada\\
$^{4}$International Centre for Radio Astronomy Research $-$ Curtin University, GPO Box U1987, Perth, WA 6845, Australia\\
$^{5}$Department of Natural Sciences, University of Michigan-Dearborn, 4901 Evergreen Rd. Dearborn, MI, 48128\\
$^{6}$Eureka Scientific, Inc., 2452 Delmer Street, Suite 100 Oakland, CA 94602, USA\\
$^{7}$Department  of  Physics  and  Astronomy,  Michigan  State  University,  East Lansing, MI 48824\\
}

\date{Accepted XXX. Received YYY; in original form ZZZ}

\pubyear{2021}

\begin{document}
\label{firstpage}
\pagerange{\pageref{firstpage}--\pageref{lastpage}}
\maketitle

\nocite{*}

\begin{abstract}
Globular clusters (GCs) have been at the heart of many longstanding questions in many sub-fields of astronomy and, as such, systematic identification of GCs in external galaxies has immense impacts. In this study, we take advantage of M87's well-studied GC system to implement supervised machine learning (ML) classification algorithms --- specifically random forest and neural networks --- to identify GCs from foreground stars and background galaxies using ground-based photometry from the Canada-France-Hawai'i Telescope (CFHT). We compare these two ML classification methods to studies of ``human-selected'' GCs and find that the best performing random forest model can reselect 61.2\% $\pm$ 8.0\% of GCs selected from HST data (ACSVCS) and the best performing neural network model reselects 95.0\% $\pm$ 3.4\%. When compared to human-classified GCs and contaminants selected from CFHT data --- independent of our training data --- the best performing random forest model can correctly classify 91.0\% $\pm$ 1.2\% and the best performing neural network model can correctly classify 57.3\% $\pm$ 1.1\%. ML methods in astronomy have been receiving much interest as Vera C. Rubin Observatory prepares for first light. The observables in this study are selected to be directly comparable to early Rubin Observatory data and the prospects for running ML algorithms on the upcoming dataset yields promising results.
\end{abstract}




\section{Introduction} \label{s:intro}
Globular clusters (GCs) are home to hundreds of thousands of gravitationally bound stars and have garnered much interest for the intriguing dynamical occurrences that are sourced within them. GCs are widely found to have a bimodal colour distribution, forming red and blue GC populations \citep{kundu2001,Harris2006,Peng2006,Brodie2012}, as well as a common bi/multimodal metallicity distribution, typically a spacially extended metal-poor population and a spatially concentrated metal-rich population \citep{Ashman1992,Brodie2012}. Both GC formation and GC properties are, however, still a constantly evolving subject of research both observationally \citep[e.g.,][]{Forbes2017,Lee2019,Usher2019,Fahrion2020} and through simulations and modelling \citep[e.g.,][]{Bastian2020,Elbadry2019,Reinacampos2019, ReinaCampos22}.

A number of black holes (BHs) and BH candidates have been uncovered in globular clusters, mostly within the Milky Way \citep[e.g.,][]{maccarone2007, strader2012, millerjones2015, giesers2018, giesers2019}. Recent simulations by \cite{Weatherford2020} show that many GCs are home to stellar-mass BHs and that these BHs may also be instrumental in the GCs' evolution and morphology. Theoretical work such as \cite{rodriguez2016} points towards globular clusters as the birthplace of BH-BH binary mergers such as those detected by the Laser Interferometer Gravitational-Wave Observatory (LIGO). GCs are also known to host a variety of X-ray sources and radio transients. These include ultra-luminous X-ray sources (ULXs) which undergo some of the most extreme mass transfer rates, and may also be indicators of BH candidates in extragalactic globular clusters \citep[and references therein]{Dage2021}. X-ray sources have also been traced back to ultracompact dwarf (UCD) hosts \citep[e.g.,][]{seth2014, Pandya2016}, of which there are many theories regarding their definition, make-up, and origin --- one common theory being that UCDs are primarily the nuclei of tidally stripped dwarf galaxies \citep{Zhang2015}. The first fast radio burst (FRB) localized to a GC was recently reported in the M81 galactic system, though the origin of this luminous flash is still unclear (\citealp{bhardwaj2021, Kirsten2021}; \citealp[see also discussion in][]{bhandari2021}). Finally, many studies suggest that globular clusters may be a potential hiding spot for the elusive intermediate mass black hole, and that the key to identifying these is studying large numbers of GCs \citep[and many references therein]{wrobel2018}. Thus, wide-scale, systematic identification of globular clusters can provide important benefits to many fields of astronomy, including high energy astrophysics and beyond \citep[e.g.,][among others]{reinacampos2021}.\\

While galactic GCs appear as a cluster of distinct point sources, extragalactic GCs can often be observed as a single point source. As such, the main issue with extragalactic GC identification is that GCs can be difficult to classify because they are easily mistaken for (1) foreground stars between the galaxy of interest and the observer or (2) distant background galaxies that appear to be associated with the observed galaxy. Extragalactic globular clusters have been identified either through ground-based photometric campaigns, spectroscopic studies, or by combining photometry with the excellent spatial resolution of the Hubble Space Telescope (HST) (e.g., \citealt{kundu2001, rhode2007,Harris2009, Jordan2009, Usher2019}).

Ground-based photometric surveys are an excellent means to create substantial GC candidate lists, but require spectroscopic follow-up to confirm the cluster nature of the candidate. HST's spatial resolution enables an estimate of the GC half light radius, which can be folded in with the photometric properties for an increased accuracy in these GC candidate lists. However, HST observations are often pointed at the centres of galaxies and many systems have not been targeted, so combing this archive yields an incomplete sample (e.g., Thygesen et al. in prep). Spectroscopy is the best means to confirm or rule out the star cluster nature of extragalactic point sources by measuring velocity dispersions \citep[e.g., the methods in][]{Illingworth1976,Ho1996, Dubath1997}. This tactic, however, can often be very telescope time intensive, as well as less sensitive to GCs close to the galaxy centres, which may be obscured by their host galaxy.

In the new era of large surveys, these methods become increasingly difficult to implement. A recent strategy for astrophysical source identification is using machine learning (ML) as a classification tool. While ML algorithms are nowhere near as accurate as spectroscopy, once trained they are quick and easy to use and, if trained properly, will still generate robust candidate lists. Such strategies have been implemented for classification of galaxies \citep[e.g.,][who used a combination of neural networks and image analysis to classify galaxies morphologically and \citealt{Ball2006} who used Sloan Digital Sky Survey data and decision trees to classify sources as galaxies, stars, or neither]{Calleja2004} and X-ray point sources \citep[e.g.,][]{Arnason2020,Pattnaik2021,Zhang2021,Tranin2021,Mountrichas2017}. These methods have also been used to successfully identify star clusters: \cite{Perez2021} and \cite{Thilker2022} used image data and image processing techniques alongside ML algorithms (in both cases convolutional neural networks) and \cite{Saifollahi2021} processed imaging data to run a K-nearest neighbours (KNN) model on only photometric colour combinations.

In this paper we employ optical, ground-based photometry to train and run supervised ML classification models to identify GC sources without spectroscopy. We train our models on sources from within M87, as it is one of the most thoroughly studied galaxies besides our own, which provides us with large catalogues of pre-classified data. Anticipating the completion of Vera C. Rubin Observatory and the resulting deep, wide data observations from the Legacy Survey of Space and Time (LSST), our aim is to create ML data classification tools that will run on these data products.\\

In Section~\ref{s:datamodels} we discuss the data and data preparation of both the training data (Section~\ref{s:trainingtest}) and the data used for external verification with human-selected GCs (Section~\ref{s:human}), as well as the model type and architecture used (Section~\ref{s:models}). Section~\ref{s:results} gives a detailed account of the results of each of our different models (Sections~\ref{s:resultrf} and \ref{s:resultnn}) and analyses the complications with separating UCDs and GCs (Section~\ref{s:ucdgc}). In Section~\ref{s:discussion} we discuss our findings, the limitations and advantages of our approach, and our recommendations for future use. Finally, we conclude our findings in Section~\ref{s:conclusion}.


\section{Data and Models} \label{s:datamodels}

To robustly train an accurate ML model that will identify extragalactic GC candidates, we require a large, uniform photometric sample. We select the photometric survey data from the Next Generation Virgo Cluster Survey (NGVS) \citep{Ferrarese2012} from observations taken by the MegaPrime instrument on the Canada France Hawaii Telescope (CFHT) using the Canadian Astronomy Data Centre (CADC) MegaPipe pipeline \citep{Gwyn2008}. This catalogue contains over 45 observations in five magnitude bands ($ugriz$), for a total of 225 merged observations (Table~\ref{t:ngvsobs} in \hyperref[s:appendixA]{Appendix A}). For each source the final combined catalogue consists of five magnitude values and five flux radius values, one for each band, as well as RA and Dec coordinates of each source's barycentre. The flux radius is defined in the NGVS catalogue documentation as the fraction-of-light radius, or half-light radius, in pixels and is included in our chosen parameters since we expect some sources (namely background galaxies) to be less resolved than others.

We train two types of ML models on classified sources for use on photometric datasets (i.e., catalogues of photometric measurements) and follow this by running independent verifications with two separate sets of human-selected GC candidates. The first human-selected catalogue is built from HST observations near the centre of M87 and consists of GC candidates with varying likelihood of being a GC, as calculated using model-based clustering methods and catalogues of expected contaminants, from \cite{Jordan2009}. The second is, like our training data, built from NGVS observations and consists of both GC candidates and other interloping sources, also with varying likelihoods of classifications (in this case calculated using colour cuts) from \cite{Oldham2016}.

 
\subsection{Training Data} \label{s:trainingtest}
To generate the training data, our formatted and unfiltered catalogue of NGVS data which contained 719,949 sources was cross-matched within 1.0 arcsec with the coordinates of each of our four classes of sources (i.e., UCDs, GCs, background galaxies, and foreground stars). Before each class of data was cross-matched with the full NGVS catalogue, they were cross-matched within 1.0 arcsec of each other to avoid overlap. Duplicate sources were left in only one of the class catalogues, with a preference in the order of UCDs, GCs, background galaxies, then foreground stars. This hierarchy was chosen due to the very limited number of our most important sources and the abundance of our less important sources. The two instances where this hierarchy was employed include: (1) sources that were listed as GCs and stars were kept in the GC dataset since there are fewer GC sources than star sources and (2) sources that were listed as UCDs and GCs were kept in the UCD dataset since the number of known UCD sources is extremely low compared to the three other class datasets.

The coordinates of UCDs used were obtained by merging catalogues from \cite{Zhang2015} and \cite{Pandya2016} for a resulting 402 sources (not all within the area of M87). The \cite{Zhang2015} catalogue is one of the largest spectroscopic datasets of UCDs, but since there are so few confirmed UCD sources in general, we chose to also use the \cite{Pandya2016} catalogue to supplement it. After cross-matching these coordinates with the NGVS data, we had a resulting dataset of 83 UCDs. The coordinates of the GCs were obtained from Strader et al. (in prep.) and offered a total of 1428 sources before cross-matching and 1188 sources after. The Strader et al. (in prep.) catalogue was chosen because it is one of the most extended specroscopic datasets of GCs in the M87 region. Note that there is some observational selection within this GC sample against sources at fainter magnitudes for two main reasons: (1) spectroscopy is typically run on only the brightest sources (which are bright either because they are close, or because they are intrinsically luminous) and (2) NGVS is only sensitive to magnitudes below 25.9 mag in the $g$ band \citep{Ferrarese2012}, making it difficult to probe the lower end of the M87 GC population.

\begin{figure}
    \centering
    \includegraphics[scale=0.35]{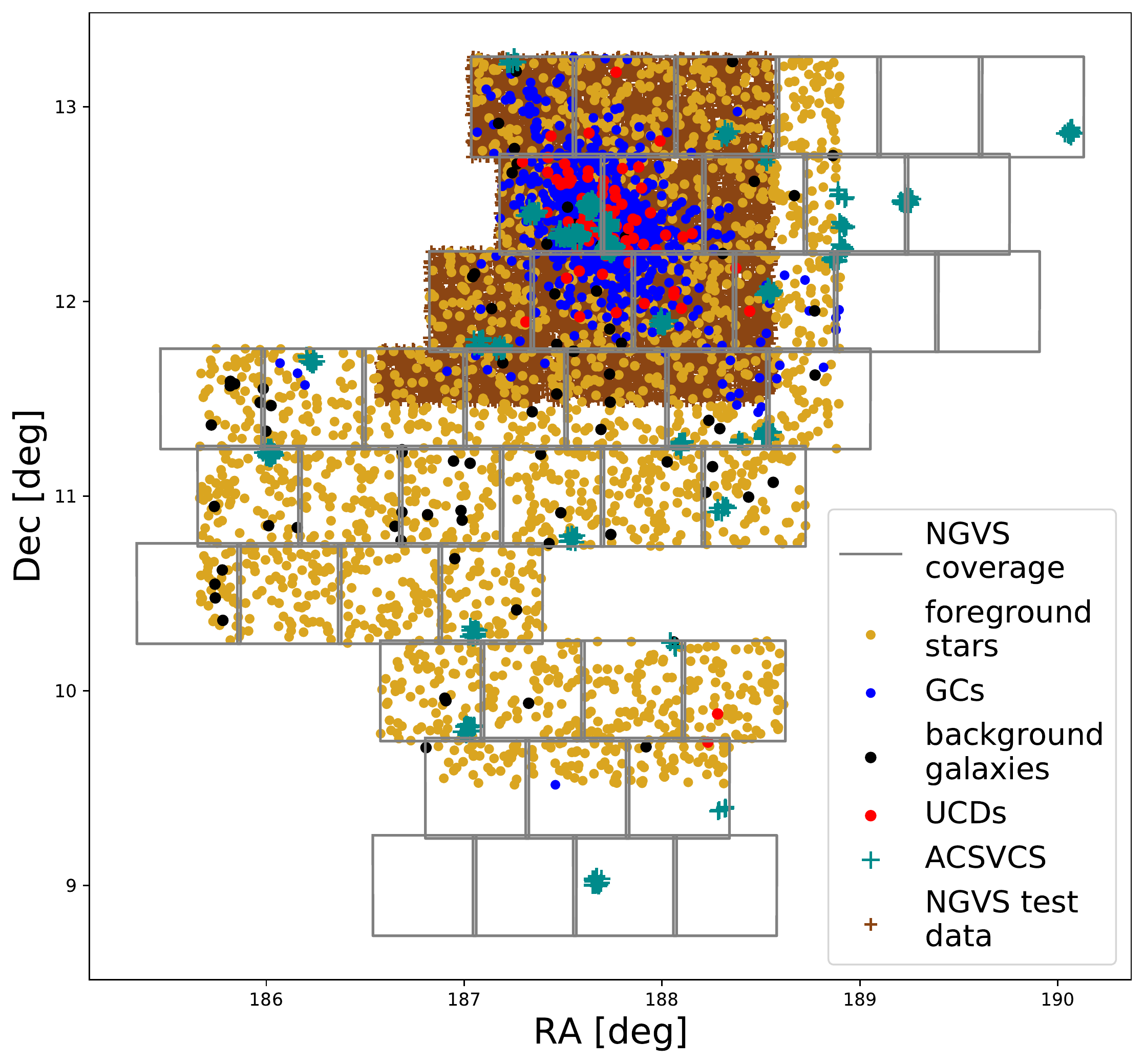}
    \caption{The spatial distribution of known sources in each of the four classes (UCDs in red, GCs in blue, background galaxies in black, and foreground stars in yellow) that make up the training dataset. The distribution of the two datasets of human-selected GC candidates are also included, as shown by the cyan and brown plus signs. Note that the ACSVCS GCs are found in small clustered groups, giving the impression of fewer sources than there are. This data was cross-matched to within 1.0 arcsec of NGVS data which covers M87 as shown by the grey boxes (Table~\ref{t:ngvsobs} in \hyperref[s:appendixA]{Appendix A}).}
    \label{f:train}
\end{figure}

\begin{figure*}
    \centering
    \includegraphics[scale=0.5]{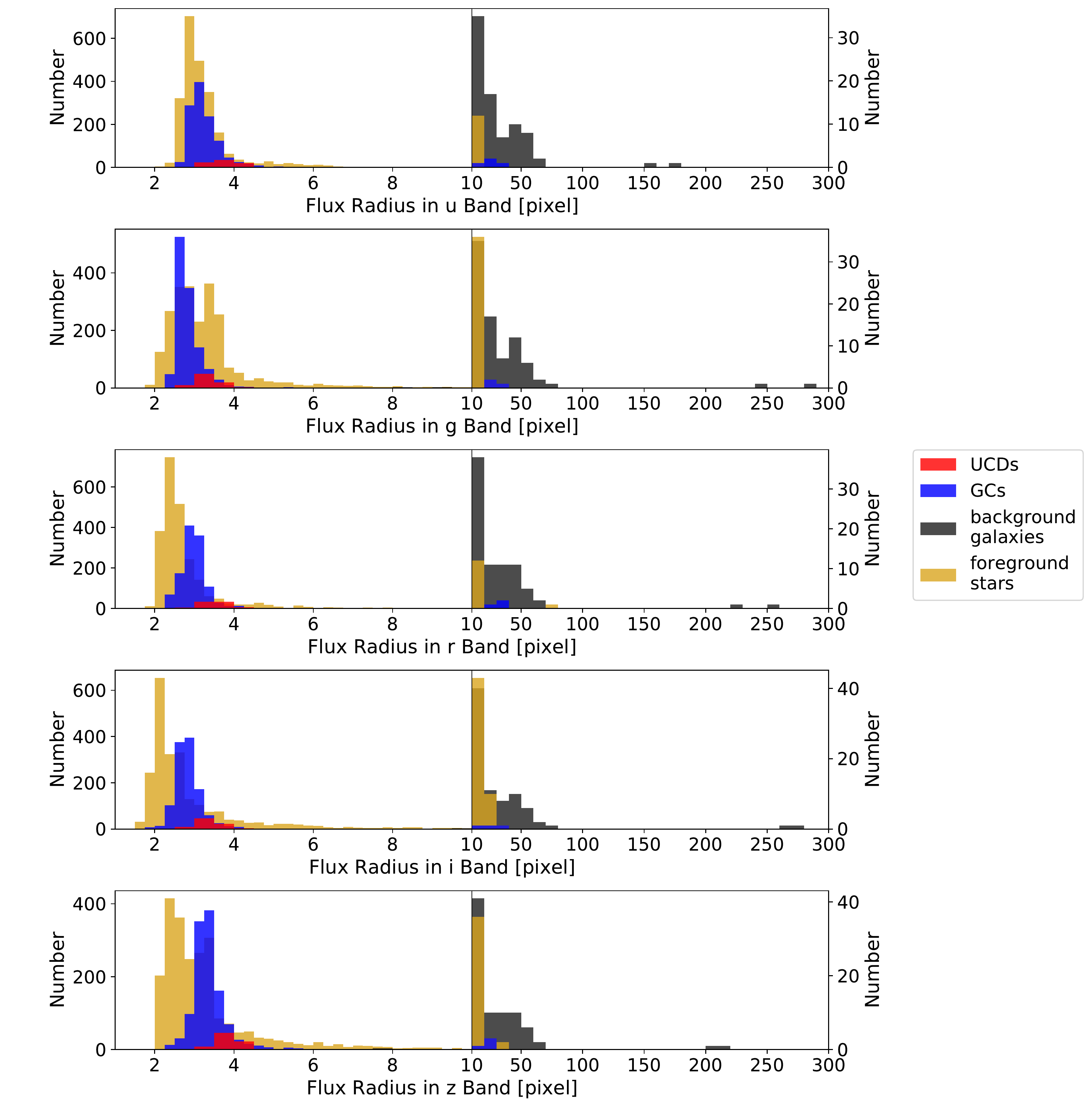}
    \caption{The flux radius distribution of of our training data with UCDs, GCs, background galaxies, and foreground stars are respectively shown in red, blue, black, and yellow in each of the five filter bands. The black vertical line in each plot indicates a change in scaling.}
    \label{f:flux}
\end{figure*}

The coordinates of background galaxies and foreground stars were both obtained by querying from various databases on NOIRLab’s Astro Data Lab\footnote{\url{https://datalab.noirlab.edu}}. The background galaxy sources were selected from the 2MASS extended source catalogue (\texttt{twomass.xsc}), which is a robust dataset consisting of non-stellar sources. These sources were queried from within the approximate coordinate area of our GC+UCD sources (to ensure that the area of sources our models are being run on has appropriate representation of all source classes) and with the flag \texttt{vc} = 1, indicating that a source was visually classified as a galaxy \citep{Skrutskie2006}. This resulted in a total of 234 background galaxy sources which was reduced to 93 sources by cross-matching their coordinates with NGVS data. The majority of the unmatched background sources are found within the M87 region but outside of the area covered by NGVS observations (i.e., the top-left region of Fig.~\ref{f:train}). The foreground star sources were downloaded from a join of two tables, the Legacy Survey's DR8 photometry catalogue of the southern region \citep[\texttt{ls\_dr8.tractor\_s}; ][]{Dey2019} and the Gaia DR2 catalogue \citep[\texttt{gaia\_dr2.gaia\_source};][]{Gaia2016,Gaia2018}, again from within our GC+UCD areas. The Gaia DR2 catalogue is large and contains a flag indicating whether a given source is a point source and values indicating the proper motion of a given source. Both of these parameters allow for reliable querying of stars. Additional constraints include a signal-to-noise ratio in $g$, $r$, and $z$ of greater than 50; a \texttt{gaia\_pointsource} value of 1, indicating the source is indeed a point source; a $z$ magnitude of greater than 10; and proper motion values in RA and Dec with a 4 sigma threshold. This resulted in an original sample of 7666 foreground star sources and was culled to 4433 through cross-matching with NGVS data. These two sources provided motivation for using flux radii as a feature (Fig.~\ref{f:flux}). While both UCDs and GCs are typically appear small and more resolved, some of our star sources appear less resolved due to their close proximity and the majority of our background galaxy sources are much less resolved due to their wider dispersions.

\begin{figure}
    \centering
    \includegraphics[scale=0.38]{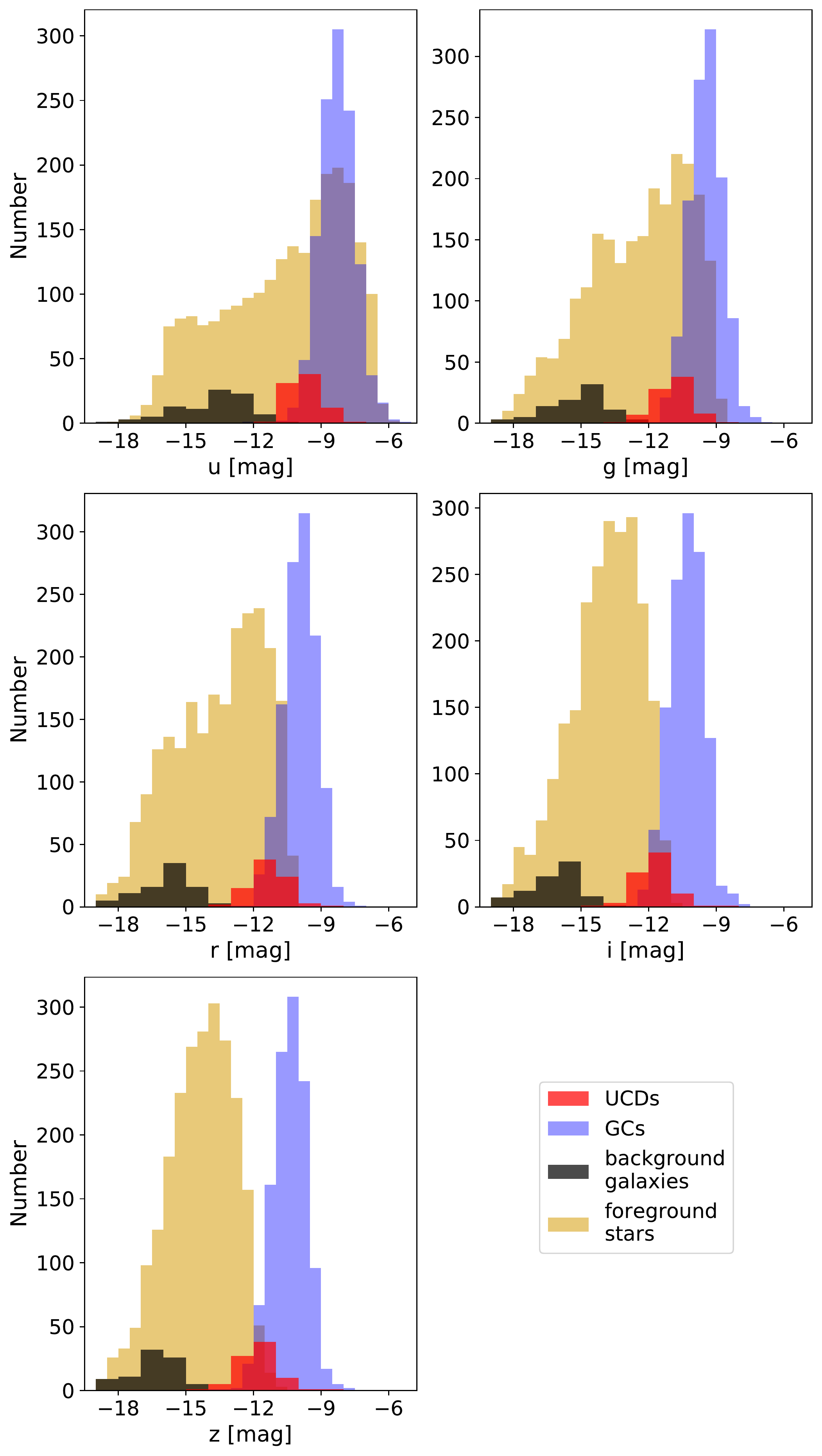}
    \caption{The absolute magnitude distribution of of our training data with UCDs, GCs, background galaxies, and foreground stars shown in red, blue, black, and yellow in each of the five filter bands. The distance to M87, 16.5 Mpc \citep{Strader2011}, was assumed for all sources in calculating absolute magnitudes.}
    \label{f:mags}
\end{figure}

\begin{figure}
    \centering
    \includegraphics[scale=0.54]{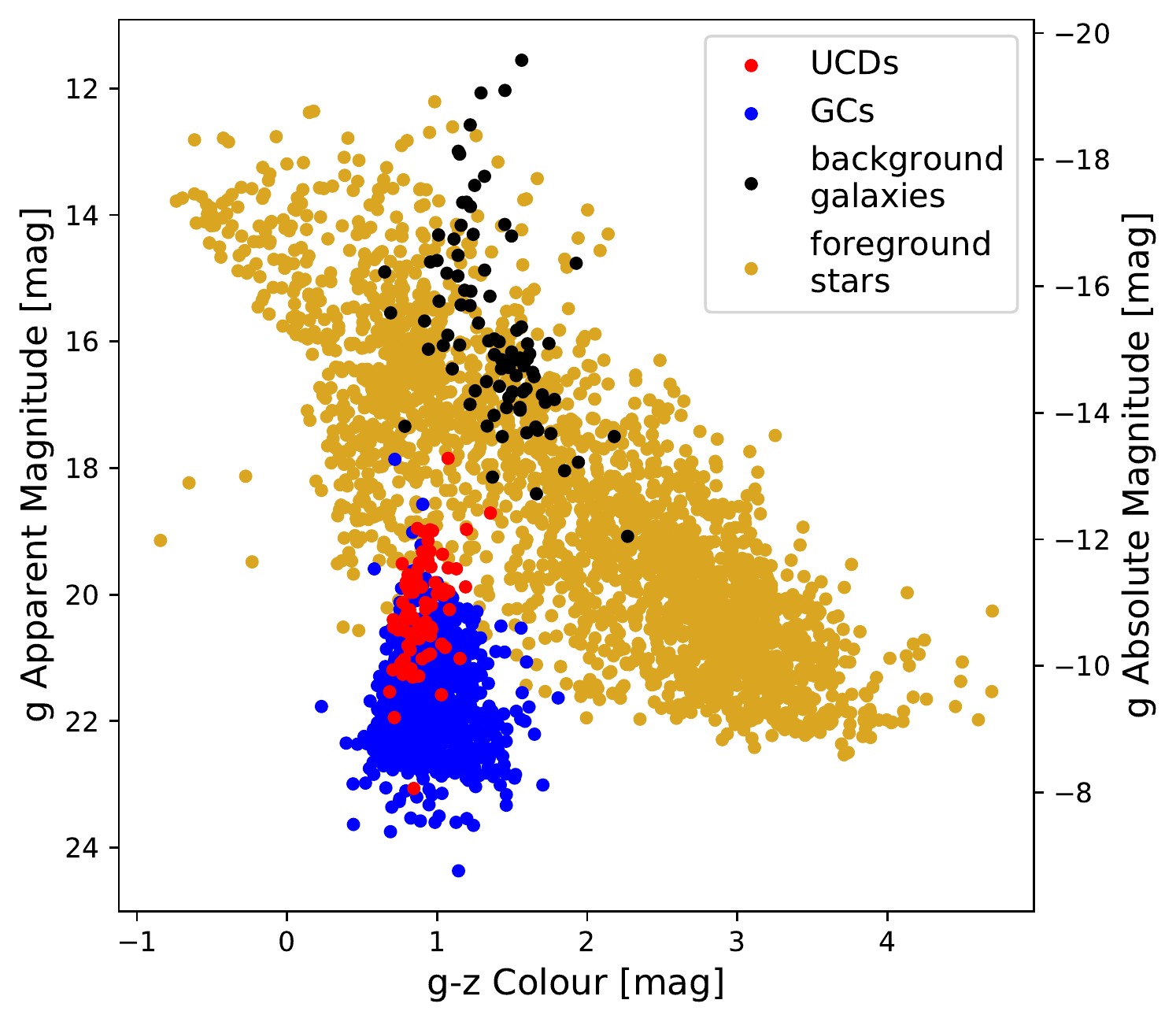}
    \caption{The \textit{g-z}/$g$ colour-magnitude diagram of each source class comprising the training dataset. (Symbols as in Fig.~\ref{f:train}) The UCD dataset is sourced from the \citet{Zhang2015} and \citet{Pandya2016} catalogues, the GC dataset from the Strader et al. (in prep.) catalogue, the background galaxy dataset from the 2MASS extended source catalogue, and the foreground star dataset from the Legacy Survey's DR8 photometry catalogue of the southern region and the Gaia DR2 catalogue (Section~\ref{s:trainingtest}).}
    \label{f:colmag}
\end{figure}

Once loaded into the ML model programs, columns were added to each dataset indicating four colours (\textit{u-g, g-r, g-i,} and \textit{g-z}) and all magnitudes were converted to absolute magnitudes assuming a distance of 16.5 Mpc, the distance to M87 \citep{Strader2011}, for all sources (Fig.~\ref{f:mags}). The use of absolute magnitudes makes it easier to run the models on other galaxies, as they would not need to be recalibrated for each new galaxy, but the models' use should still be limited to galaxies of similar distances since the relationship between the features of each source will change with distance. To ensure the use of only the most reliable data and eliminate any sources listed with unphysical magnitudes, which may indicate missing or erroneous data, the datasets were then trimmed to ensure magnitudes in all bands fall within the the nominal limiting magnitude of each observation and band. One additional reduction was applied to the stars dataset to ensure that only the stars with the highest proper motion (i.e., the sources that are the brightest and fastest and therefore the most likely to be real stars) were used. This was done by using a generous trim of any data with proper motion in RA and/or Dec between -10 mas/year and 10 mas/year. An additional reduction was also applied to the GC dataset regarding radial velocity. M87 has a heliocentric radial velocity of 1284 km/s \citep{Cappellari2011}, so only GCs with velocities in the range of 200-2400 km/s were used to avoid any potential confusion between low velocity GCs and foreground stars. The multiple cross-matches resulted in many columns in each dataset, but only magnitude (in each of the 5 $ugriz$ bands), colour (each of the four aforementioned combinations), and flux radius values (in each of the 5 $ugriz$ bands) were used as input features for the models to run on (Table~\ref{t:features}). After removing outliers and filtering the GCs and stars, our remaining datasets contained 83 UCDs, 1160 GCs, 90 background galaxies, and 2346 foreground stars (Fig.~\ref{f:train}, Table~\ref{t:datacounts}). These populations have distinct source properties, but overlap in colour-magnitude space as seen in Fig.~\ref{f:colmag}.

\subsection{Human-selected GCs for external verification} \label{s:human}
The first catalogue used to verify and test our models against human-selected GC candidates was the ACS Virgo Cluster Survey (ACSVCS) catalogue of globular cluster candidates which was created based on data taken by HST \citep{Jordan2009}. We do have an inherent bias when running our models on this dataset, as the NGVS data has a limiting magnitude of 25.9 mag in the $g$ band \citep{Ferrarese2012}, whereas the ACSVCS catalogue has some magnitude values fainter than 26 mag in the $g$ band (since availabe HST data often probes fainter magnitudes than current data available from ground-based telescopes like CFHT). The ACSVCS catalogue does not include magnitude values in the \textit{u, i,} and $z$ bands or flux radius values, so it is necessary to cross-match with CFHT data, preventing us from running our model on the faintest sources in the ACSVCS catalogue. As such, much like the training data, the RA and Dec coordinates of this dataset were cross-matched within 1.0 arcsec of our main NGVS catalogue (Section~\ref{s:trainingtest}). The AVSVCS dataset was also cross-matched within 1.0 arcsec of the training datasets to ensure there was no overlap. This resulted in a total of 646 sources from the original 12,763 being input into our models (Fig~\ref{f:train}, Table~\ref{t:datacounts}).

Unlike the previous datasets used to construct our training data, the ACSVCS sources are listed with values indicating the probability of each being a globular cluster (pGC), as calculated by \cite{Jordan2009} using model-based clustering methods and catalogues of expected contaminants. Once incorporated, the sources were again trimmed to be within the the nominal limiting magnitude of each observation and band to eliminate missing or erroneous data, leaving a total of 645 sources, and the apparent magnitudes were converted to absolute magnitudes. Each model (detailed in Section \ref{s:models}) except those focused on only UCDs vs GCs was then run on this dataset and we based our models’ success on how well they reselected this list of GCs. Since this dataset was created using a space telescope rather than a ground-based telescope, it includes partially resolved GCs and serves as a robust and well selected list of GC candidates. Reselecting a large number of these sources is essential for any well-performing model. The only drawback to this diagnostic is that the catalogue does not include any background or foreground contaminants, meaning that running models on this data gives no information on how well a given model can recognize background and foreground sources.

The second catalogue used to verify and test our models was the \cite{Oldham2016} catalogue of 17,620 GC candidates, which were chosen by colour cuts and, like our training dataset, consisted of NGVS photometric data taken by the CFHT MegaPrime instrument. These sources were cross-matched within 1.0 arcsec of our prepared NGVS catalogue and the same trimming and conversion of magnitudes were done. This dataset, which we refer to as the NGVS test set, was then also cross-matched within 1.0 arcsec of the training datasets to ensure there was no overlap. This resulted in 11,862 sources after cross-matching and 11,860 sources after trimming (Fig~\ref{f:train}, Table~\ref{t:datacounts}). A key difference between this catalogue and the last is that this dataset includes interloping contaminant sources and has more specific classification probability values. The latter were obtained using functions of galactocentric radius, magnitude, and colour, and include the probability of each source belonging to the red GC population (labelled as "pRed" in their catalogue), the blue GC population (pBlue), the Milky Way/Sagittarius stellar interloper population (pMW), or the uniform-colour interloper population (pInt as labelled in their paper or pStar as labelled in their corresponding catalogue). The latter interloper population may consist of other contaminants such as background galaxies or sources within the Virgo overdensity. Before running our models on this catalogue, we summed the blue and red GC probabilities to get one value indicating the probability of being a star cluster (pC) and summed the interloper probabilities to get a second value indicating the probability of being a contaminant (pNC). These two combinations were summations rather than averages so that the sum of all possible probabilities for a given source (i.e., pC and pNC) would equal 1. This pNC parameter allowed for an additional verification of not only how well our models can reselect GCs+UCDs, but also how well they can reselect background and foreground sources.

\begin{table}
    \centering
    \caption{Final count of sources on which our models were run of each class in the training dataset and in both human-selected catalogues.}
    \begin{tabular}{ c|cc|cc }
        \hline
        catalogue & UCD & GC & \makecell{background \\ galaxy} & \makecell{foreground\\star}\\ 
        \hline
        NGVS training set & 83 & 1160 & 90 & 2346\\
        ACSVCS & \multicolumn{2}{c|}{645} & 0 & 0\\
        NGVS test set & \multicolumn{2}{c|}{6951} & \multicolumn{2}{c}{4909}\\
        \hline
    \end{tabular}
    \label{t:datacounts}
\end{table}


\subsection{Models} \label{s:models}
Machine learning can be implemented in many different ways with a variety of different types of training algorithms. We aim to compare the performance of two popular approaches ideal for supervised classification tasks: random forest (RF; Section~\ref{s:rf}) and deep neural networks (NN; Section~\ref{s:nn}). The main motivation behind the choice of these two ML algorithms is that we aim to make our models accessible for future users. NN and RF are two of the more common algorithms for supervised classification based tasks and have a minimal barrier to entry for ML beginners. For both approaches, three separate models were trained and tested: (i) a binary classification model of GCs+UCDs and contaminant sources (i.e., foreground stars and background galaxies), (ii) a multi-class classification model of GCs+UCDs, foreground stars, and background galaxies, and (iii) an auxiliary binary classification model of only GCs and UCDs. The first two types of models were trained and tested on our full training and testing datasets, but the third model was only run on UCD and GC sources, since these are the only types of objects it can classify. There is some disagreement on what precisely constitutes a UCD and how precisely it differs from a GC, but we nonetheless aimed to see how well this auxiliary model performed based on the UCD catalogue we had access to and under the assumption that all sources in it are in fact UCDs.

\begin{table}
    \centering
    \caption{All models utilized all 14 of the below features, except for NN3 which utilized only 9, the colour features and the flux radius features.}
    \begin{tabular}{ ccc }
        \hline
        feature & magnitude bands & feature count\\ 
        \hline
        magnitude & \textit{u, g, r, i, z} & 5\\
        colour & \textit{u-g, g-r, g-i, g-z} & 4\\
        flux radius & \textit{u, g, r, i, z} & 5\\
        \hline
    \end{tabular}
    \label{t:features}
\end{table}


\begin{table*}
    \centering
    \caption{Hyperparameters used in each random forest and neural network model. The RF hyperparameters indicate the number of decision trees to be built, the metric formula used to make decisions at each tree node, and the minimum number of sources that should end up at each leaf. The NN hyperparameters indicate how strongly the model should correct itself, how much data it should be run on before updating, and how many epochs the training process should last.}
    \begin{tabular}{ cccc||cccc }
        \hline
        RF hyperparameter & RF1 & RF2 & RF3 & NN hyperparameter & NN1 & NN2 & NN3\\ 
        \hline
        n\_estimators & 500 & 500 & 500 & learning rate & 0.001 & 0.001 & 0.01\\
        criterion & entropy & entropy & gini & epochs & 80 & 100 & 90\\
        min\_samples\_leaf & 5 & 1 & 1 & batch size & 80 & 80 & 10\\
        \hline
    \end{tabular}
    \label{t:hparams}
\end{table*}

\subsubsection{Random forest approach} \label{s:rf}
RF is a classification algorithm which utilises many decision trees to classify objects. Decision trees are tree-like structures similar to flowcharts with decision nodes determining which branches to send the object down, then branching off further to other nodes. At the end of the trees are leaves which represent the possible classification outcomes (class labels) and serve as the output. Individual decision trees are susceptible to overfitting and are not flexible to changes in data. The random forest classification helps to correct these disadvantages by creating many decision trees which each use random subsets of the input features and then averaging over them. The use of random subsets and many trees allows the trees to be uncorrelated and reduces overfitting. 

Our RF algorithm was implemented using \texttt{RandomForestClassifier} from \texttt{sklearn}\footnote{\url{https://scikit-learn.org}} \citep{scikit-learn}. Three separate classifiers were created for different classification purposes. The first model (RF1) is a binary classification model distinguishing between GCs+UCDs and contaminant sources (i.e., foreground stars and background galaxies). The second model (RF2) is a 3-class classification model distinguishing between GCs, background galaxies and foreground stars. Lastly, the third model (RF3) is another binary model which was created to distinguish between GCs and UCDs and, as such, was only trained on pre-classified UCD and GC sources rather than the full training dataset. All three models utilised the same training data divided with an 80-20\% data split where all the sources of stars, background galaxies, and UCDs were used, but a trimmed subset of only 600 GCs were used to minimize class imbalance. All three models also utilized all 14 of the features available (Table~\ref{t:features}) and had hyperparameters optimized using \texttt{sklearn}'s \texttt{GridSearchCV} \citep{scikit-learn} which are listed in Table~\ref{t:hparams} and include values indicating the number of decision trees to be built, the metric formula used to make decisions at each tree node, and the minimum number of sources that should end up at each leaf. Aside from the hyperparameters listed in Table~\ref{t:hparams}, all other hyperparameters were left at their default values with the exception of \texttt{class\_weight} which was set to "\texttt{balanced\_subsample}" to alleviate the imbalance in class sizes for all models.


\subsubsection{Neural network approach} \label{s:nn}
Another common implementation of machine learning is the use of NN. NN models are a collection of nodes (also referred to as neurons) organized into one or more layers. Through supervised machine learning, a model is trained by passing preprocessed data into the algorithm, which continually updates its nodes and the weights between nodes on different layers as it is given more data. When training a model it is necessary to split the training data into what is typically termed a training set and a test set to assess metrics before releasing the model on new, unclassified data. A typical data split is 80\% in the training set and 20\% in the test set, which is the split used in all models in this paper. A second split of the data is made within the training set and is termed the validation set (in this case a 70\%-30\% split was used). This set is used throughout training to test how each update of nodes and weights fared and how much it should update itself on the next run.

The training data consists of a variety of features, or parameters, along with either a class or numerical label, all of which is passed into the algorithm. Additionally, a set of initial parameters, the hyperparameters, are hardcoded into the training algorithm by the user and can be tweaked to improve model performance. These hyperparameters include values indicating how drastically the model should update itself, how much data it should be run on before updating, and how many epochs this process should last. The hyperparameters used for our NN models were chosen for best performance by trial and error and are detailed in Table~\ref{t:hparams}. Once a model has been trained on these feature-label pairings, it can then predict either a classification or a regression (depending on the type of NN model) for new data that is passed in with the same features as the training data. Considering our goal is source classification, it is most appropriate to have our models predict a class.

Three NN models were created for our dataset, each with a different number of nodes and layers. The first model (NN1) is a binary classification model that was trained on GCs+UCDs and contaminant sources (ie. foreground stars and background galaxies) with two layers of 10 nodes each. The second model (NN2) is a 3-class classification model and was trained on GCs+UCDs, foreground stars, and background galaxies with two layers of 20 and 10 nodes. The third and final model (NN3) is another binary classification model which was trained only on GCs and UCDs with two layers of 50 and 30 nodes. For this model it was necessary to trim the number of GCs it was trained on from 1160 sources to 250 due to the small number of UCD sources available. Had this class not been trimmed there would be a significant class imbalance and the model would not function properly. Typically with class imbalance data, models have a tendency to assume all sources belong to the class with the larger dataset, since that would technically make the model more accurate, despite not learning anything about the other, smaller classes.

The first two models NN1 and NN2 were both trained on 14 features (Table~\ref{t:features}), including absolute magnitude (\textit{u, g, r, i,} and $z$), colour (\textit{u-g, g-r, g-i,} and \textit{g-z}), and flux radius values for each of the five magnitude bands, whereas the third model NN3 was trained only on colour and flux radius values. When tested by trial and error, NN3 performed best when only run on these 9 features, whereas NN1 and NN2 performed best on all available features. One possible reason for this could be that NN3 was trained on significantly fewer sources than the other two NN models. A larger number of features without a proportionately large number of sources may have resulted in too much noise for the model to infer a clear idea of what constitutes each class. Note that each model should only be used on data that it is expected to classify. For instance, the second binary model NN3 should not be used on an unfiltered catalogue as it will not know how to deal with anything other than GCs or UCDs, but will perform as expected on a catalogue of sources believed to contain only those types of sources. Additionally, each model must only be run on data which has all of the features that it was trained on, as it will not execute on data that is missing features.


\section{Results} \label{s:results}
The three RF and three NN models were each trained on 80\% of the training data prepared as detailed in Section~\ref{s:trainingtest} before then being tested on the remaining 20\% of the training data. Multiple metrics were used to get a clear view of how the models performed on this portion of the training data, especially in cases where there was a class imbalance in the data (e.g., in RF2/NN2 and RF3/NN3 due to a low number of confirmed sources of background galaxies and UCDs, respectively). These metrics include overall accuracy across all classes (i.e., the fraction of all classifications that are correct), precision of each individual class (i.e., the fraction of a given predictive class' identifications that are correct, which gives a measure of false positives), recall of each individual class (i.e., the fraction of a given class' actual sources that were correctly classified, which gives a measure of false negatives), and comparisons with human-selected GCs from the ACSVCS and NGVS test catalogues. Considering each of these metrics rather than just one or some provides a clearer idea of model performance, however, when training, more importance was placed on the minimization of false positives (i.e., maximizing precision) than the minimization of false negatives (i.e., maximizing recall), since a more reliable, but incomplete candidate list is preferable to a more complete, but unreliable list in this case. The aim of this paper is not simply to find all likely candidates for targeted follow-up, but to do so with a focus on data reduction and accuracy.

\begin{figure}
    \centering
    \includegraphics[width=1.67in]{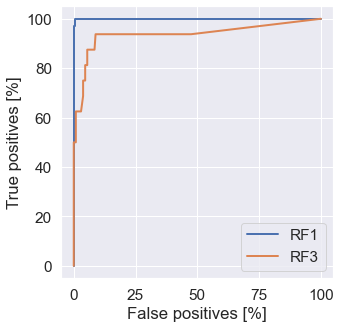}
    \includegraphics[width=1.67in]{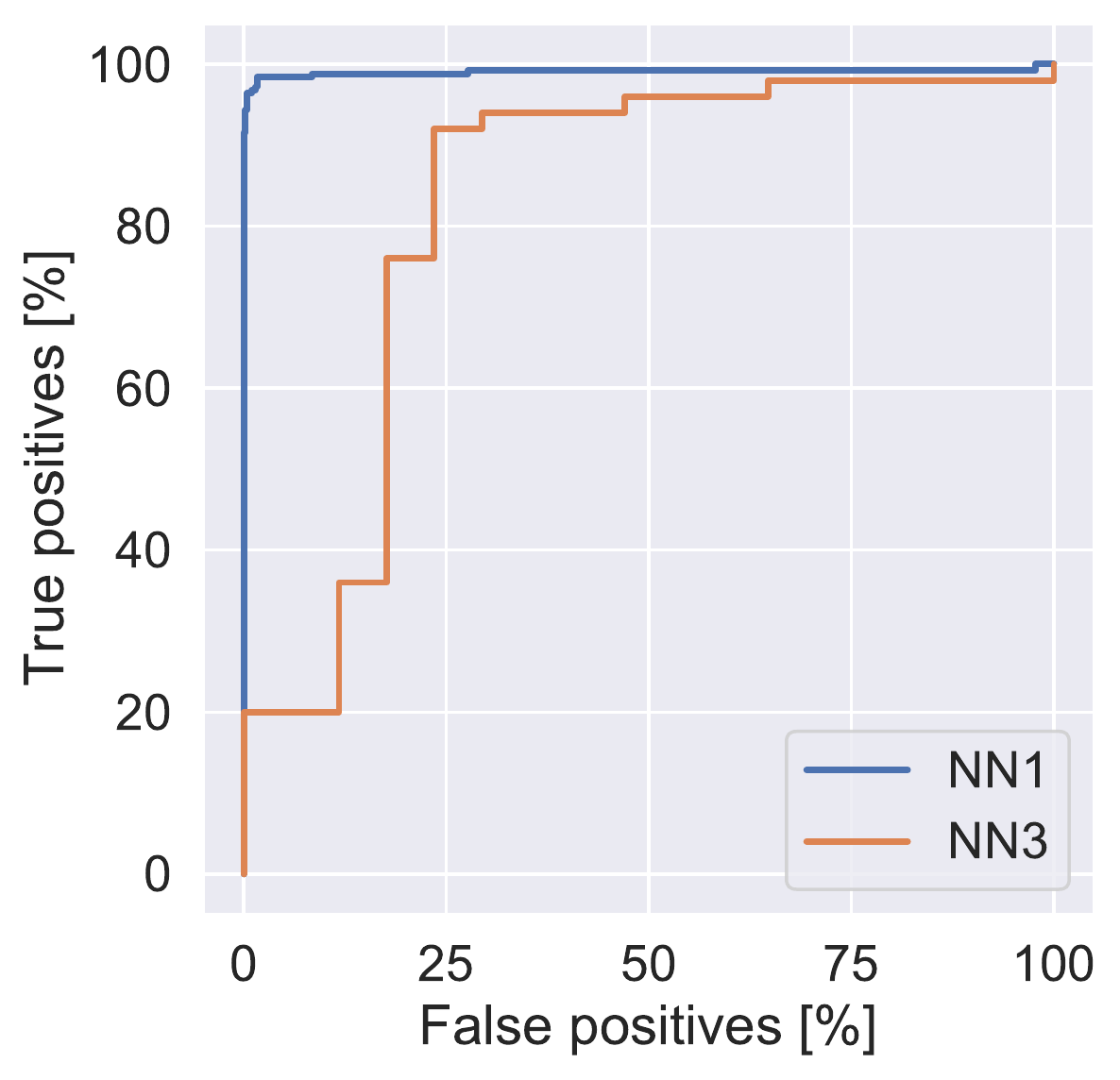}
    \caption{Receiver operating characteristic curves (ROC curves) of both RF binary classification models (left) and both NN binary classification models (right), which show the rate of false positives against that of true positives. All four ROC curves have a large area under the curve (AUC) indicating high performance, but RF1 and NN1 are the better performers given their greater AUC. RF1 and RF3 have AUCs of 0.99 and 0.94 and NN1 and NN3 have AUCs of 0.99 and 0.82.}
    \label{f:roc}
\end{figure}

Since every time a model is trained it performs slightly differently, each model was trained and tested 20 times to see how each performs on average. A summary of the averages of all metrics for each RF and NN model can be found in Table~\ref{t:results}. All metric values aside from the comparisons with human-selected GCs were calculated from the confusion matrix --- the usual method for visualizing results of ML models --- of each model which details how individual sources are identified. These are structured with the predicted class labels on the x-axis and the known class labels on the y-axis. Confusion matrices for each of the 6 models can be found in Fig.~\ref{f:cm} in \hyperref[s:appendixB]{Appendix B}. We view our results through an additional visualization tool termed the receiver operating characteristic curve (ROC curve), which plots the rate of false positives against the rate of true positives in characterizing the test data, considering different classification thresholds (i.e., decision boundaries). An ideal ROC curve would have a maximal area under the curve whereas a random classifier would result in a perfect diagonal line. ROC curves are typically used on binary classifiers and, as such, the ROC curves for RF1, RF3, NN1, and NN3 are shown in Fig.~\ref{f:roc}. An easy way to better understand ROC plots is by calculating a model's area under the ROC curve (AUC). An AUC score ranges from 0 to 1, meaning an AUC of close to 1.0 would implicate a model's predictions are entirely correct, a score close to 0.0 would imply the predictions are entirely wrong, and a score close to 0.5 would imply an uninformative model. For the ROCs shown in Fig.~\ref{f:roc}, our binary RF models have AUCs of 0.99 and 0.94, respectively, and our binary NN models have AUCs of 0.99 and 0.82, respectively. While these are very high scores, an important caveat is that ROCs (and AUCs as a result) tend to be overly optimistic for class imbalance data such as ours, especially when the minority class is the the main focus of the model \citep{He2013}. This means that for NN3 and RF3 where UCDs are the main focus, but the source count is minimal, the ROC curves are expected to be especially optimistic.

\begin{table*}
    \centering
    \caption{Metrics for each of the 6 models, with precision and recall values listed for each class. The ACSVCS column indicates how well the models can reselect GCs classified using HST data and the NGVS test column indicates how well the models can reselect both GCs and contaminants classified using NGVS data. The ACSVCS catalogue was cross-matched with NGVS data and both human-selected catalogues were cross matched with our training data to avoid overlap. A detailed definition of each metric is found in Section~\ref{s:results}.}
    \begin{tabular}{ cccccc }
        \hline
        model & accuracy (\%) & precision (\%) & recall (\%) & ACSVCS & NGVS test\\ 
        \hline
        \hline
        RF1 & 99.4 $\pm$ 0.2 & \makecell{GCs+UCDs: 98.9 $\pm$ 0.6\\ contaminants: 99.6 $\pm$ 0.2} & \makecell{GCs+UCDs: 99.2 $\pm$ 0.4\\ contaminants: 99.4 $\pm$ 0.3} & 59.4 $\pm$ 3.7 & 91.0 $\pm$ 1.2\\
        \hline
        RF2 & 99.5 $\pm$ 0.3 & \makecell{GCs+UCDs: 99.3 $\pm$ 0.5\\ background galaxies: 98.4 $\pm$ 3.4\\ foreground stars: 99.7 $\pm$ 0.2} & \makecell{GCs+UCDs: 99.4 $\pm$ 0.5\\ background galaxies: 97.9 $\pm$ 4.3\\ foreground stars: 99.6 $\pm$ 0.3} & 61.2 $\pm$ 8.0 & 86.0 $\pm$ 3.8\\
        \hline
        RF3 & 95.4 $\pm$ 0.8 & \makecell{UCDs: 71.0 $\pm$ 9.9\\ GCs: 96.6 $\pm$ 0.7} & \makecell{UCDs: 47.8 $\pm$ 11.0\\ GCs: 98.6 $\pm$ 0.6} & - & -\\
        \hline
        NN1 & 98.4 $\pm$ 0.4 & \makecell{GCs+UCDs: 98.2 $\pm$ 1.0\\ contaminants: 98.5 $\pm$ 0.5} & \makecell{GCs+UCDs: 97.0 $\pm$ 1.0\\ contaminants: 99.1 $\pm$ 0.5} & 95.0 $\pm$ 3.4 & 57.3 $\pm$ 1.1\\
        \hline
        NN2 & 98.0 $\pm$ 0.7 & \makecell{GCs+UCDs: 97.8 $\pm$ 1.5\\ background galaxies: 77.6 $\pm$ 5.0\\ foreground stars: 99.0 $\pm$ 0.5} & \makecell{GCs+UCDs: 97.2 $\pm$ 1.0\\ background galaxies: 94.4 $\pm$ 0.0\\ foreground stars: 98.5 $\pm$ 0.9} & 94.8 $\pm$ 1.5 & 57.2 $\pm$ 0.9\\
        \hline
        NN3 & 81.1 $\pm$ 3.0 & \makecell{UCDs: 71.2 $\pm$ 7.3\\ GCs: 84.5 $\pm$ 5.0} & \makecell{UCDs: 47.9 $\pm$ 22.4\\ GCs: 92.4 $\pm$ 6.0} & - & -\\
        \hline
    \end{tabular}
    \label{t:results}
\end{table*}

\subsection{Effectiveness of Random Forest Approach} \label{s:resultrf}

RF1 had very high performance with an overall accuracy of 99.4\% +/- 0.2\%, GC+UCD precision of 98.9\% +/- 0.6\%, and GC+UCD recall of 99.2\% +/- 0.4\%. RF2 also had high performance, showing near identical metrics within error to RF1 with an overall accuracy of 99.5\% +/- 0.3\%, GC+UCD precision of 99.3\% +/- 0.5\%, and GC+UCD recall of 99.4\% +/- 0.5\%. RF1 and RF2 were then used to classify the ACSVCS GC candidates where the results were compared with the listed pGC values for each source. RF2 proved to be the better model at selecting GC candidates with the model reselecting 61.2\% +/- 8.0\% of the GCs and RF1 reselecting 59.4\% +/- 3.7\% of the GCs. The NGVS test set sources were used similarly to ACSVCS GC candidates to determine how well our models reselect GCs. The NGVS test set sources with pRed and/or pBlue values greater than 0.5 were treated as GCs, sources with pC (i.e., pRed + pBlue) values greater than 0.8 were also deemed as GCs, and all other sources were considered to be contaminants. Here RF1 had higher performance at reselecting GCs from the catalogue with RF1 reselecting 91.0\% +/- 1.2\% and RF2 reselecting 86.0\% +/- 3.8\% of the NGVS test sources.

RF3 has the worst performance out of the three RF models, with significantly lower performance compared to RF1 and RF2 over all metrics. This is somewhat expected as, not only are there minimal UCD sources to train on and much fewer distinguishing factors between UCDs and GCs than between other classes, there is also much less agreement on what precisely constitutes a UCD. These issues are discussed further in the following Section~\ref{s:ucdgc}. RF3 has an overall accuracy of 95.4\% +/- 0.8\%, UCD precision of 71.0\% +/- 9.9\%, and UCD recall of 47.8\% +/- 11.0\%. This model, unlike the previous two, was not run on either the ACSVCS or the NGVS testing catalogues, as neither would be an appropriate diagnostic tool. This model can only classify sources as UCDs or GCs and since the ACSVCS GC catalogue was created to intentionally exclude UCDs, running the model on this dataset would not give us any information regarding how well it can separate UCDs from GCs. Similarly, the sources in the NGVS test set include sources that are not strictly GCs or UCDs, and, as such, this model cannot be run on this catalogue since it cannot possibly classify these contaminant sources as anything other than star clusters. A comparison of the ROC curves of RF1 and RF3 are found in Fig.~\ref{f:roc}. A large area under the curve is ideal and, as such, while both perform well, RF1 is evidently the higher performer.


\subsection{Effectiveness of Neural Network Approach} \label{s:resultnn}
NN1 and NN2 had very similar performance, with many metrics that are equal within errors. This is expected since both models classify GCs+UCDs in the same way, with the only difference being how they classify contaminants. The two models respectively had an overall accuracy of 98.4\% +/- 0.4\% and 98.0\% +/- 0.7\%, GC+UCD precision of 98.2\% +/- 1.0\% and 97.8\% +/- 1.5\%, and GC+UCD recall of 97.0\% +/- 1.0\% and 97.2\% +/- 1.0\%. While NN1 metrics are slightly higher, this can be attributed to there being one less class for the model to choose from and is not a significant enough difference to claim one model as performing better than the other. Once created, these models were run on the ACSVCS GC candidates and the values predicted by the model were compared to the listed pGC values for each source. Both models, again, had very similar performance and reselected 95.0\% +/- 3.4\% and 94.8\% +/- 1.5\% of the GCs respectively. These models were then run on the NGVS testing catalogue of sources after combining pRed and pBlue values into one pC (probability of star cluster) value and pMW and pInt/pStar values into one pNC (probability of contaminant) value. We considered sources with either pRed or pBlue values above 0.5 or pC values above 0.8 to be GCs and all others sources to be contaminants. The two models then correctly reclassified 57.3\% +/- 1.1\% and 57.2\% +/- 0.9\% of the NGVS test sources. Note that while the NN models had much higher performance on the ACSVCS catalogue, the RF models had much higher performance on the NGVS test catalogue.

Compared to the other two models, NN3 has much lower performance across all metrics. This is again somewhat expected due to minimal UCD sources to train on, fewer distinguishing factors between UCDs and GCs than between other classes, and the lack of agreement on what precisely constitutes a UCD. This model has an overall accuracy of 81.1\% +/- 3.0\% and has 71.2\% +/- 7.3\% precision and 47.9\% +/- 22.4\% recall for UCDs. This indicates that while the model may miss near half of the UCDs present, when it does classify a source as a UCD it is correct near 71\% of the time. Using this model on other GC catalogues may therefore result in an incomplete but still interesting list of UCD candidates. This model, like RF3, was run on neither the ACSVCS catalogue nor the NGVS test catalogue for the reasons described in Section~\ref{s:resultrf}. A comparison of the ROC curves of NN1 and NN3 is found in Fig.~\ref{f:roc}. A large area under the curve is ideal and, as such, while both perform well, NN1 is evidently the higher performer.


\subsection{Ultra-compact Dwarfs vs. Globular Clusters} \label{s:ucdgc}
There are multiple theories regarding the definition, make-up, and origin of UCDs, but \cite{Zhang2015} found that their results support the theory that UCDs are primarily the nuclei of tidally stripped dwarf galaxies. Results from \cite{Saifollahi2021} suggest that UCDs can be distinguished from GCs, and that ML algorithms can be used (in addition to follow-up inspection) to achieve this task. However, they also recognize that this tidally stripped scenario regarding UCDs may be biased. They suggest that, as catalogues of studied UCDs are typically taken from surveys focused on high density environments, since that is where most spectroscopic surveys are done, there is very little information on UCDs or UCD-like objects on the outskirts of galaxy clusters \citep{Saifollahi2021}. More research on UCDs farther out may or may not change the generally accepted theories, but it would likely give a more thorough view of the UCD population and allow for more advanced classification algorithms. Models RF3 and NN3 were therefore run on datasets that rely heavily on a few assumptions: that UCDs are observationally distinct enough from GCs to effectively sort and separate them; that they differ from GCs as postulated in \cite{Zhang2015} as having a larger size, a possible size-luminosity relation, and a mass greater than about $2x10^6 M_{\sun}$; and that all sources listed in the \cite{Zhang2015} catalogue are in fact UCDs. 

\begin{figure}
    \centering
    \includegraphics[width=3.35in]{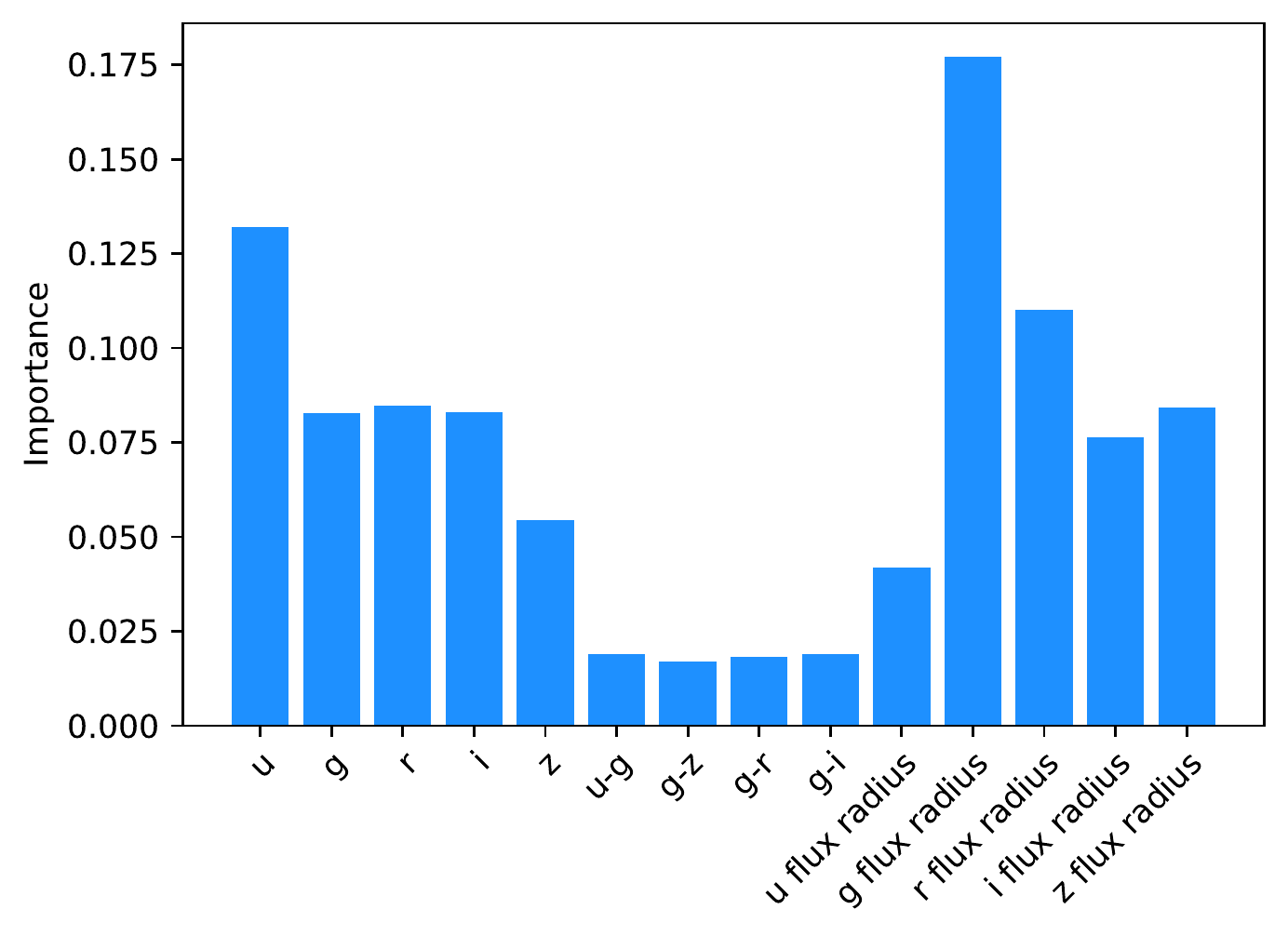}
    \caption{Impurity based feature importances of model RF3 normalized to 1. While this model does use all available features, magnitude features and flux radius features are shown to be similarly important and both more important than colour features when classifying UCD and GC sources.}
    \label{f:feat_imp}
\end{figure}

Fig.~\ref{f:feat_imp} shows the relative importance of each feature used in training RF3 as reported by the model itself. Unlike NN packages, the \texttt{RandomForestClassifier} class includes an attribute, \texttt{feature\_importances\_}, which indicates impurity-based feature importances \citep{scikit-learn} and can be used to compare the relative importance of each feature for a given model during training. Also known as the Gini importance, the importance of each feature is calculated based on how much it reduces the criterion (i.e., the function that measures the quality of each new branch in RF trees). This shows that while the model does use all available features to distinguish between sources, it places the most importance on magnitude and flux radius (i.e., size) features. This is in agreement with our above assumptions regarding UCDs. NN3, unlike RF3, performed best when only run on the colour and flux radius features. This can likely be attributed to this combination of features being the fewest possible features that still provide the model with a substantial amount of information. NN3 was trained on significantly fewer sources than the other two NN models and a larger number of features without a proportionately large number of sources may have resulted in too much noise for the model to infer a clear idea of what constitutes each class. This may not have been an issue for RF3, but RF models and NN models are structured very differently.

With our small dataset of only 83 UCD sources, we were able to build two models with mediocre performance that only looked at UCDs vs GCs and still required a very significant trim of sources to avoid a strong class imbalance in the data. This trim implies a loss of information from the unused GC sources which may affect model performance. To improve this would require either choosing only the most representative GCs to use in the training data rather than randomly selecting them or to simply have a larger UCD dataset, potentially finding and incorporating outskirt UCDs. A larger dataset would also allow for a model involving other contaminant sources as well. Models that classify UCDs, GCs, and contaminant sources are impractical with our current datasets since our small UCD dataset either becomes invisible when combined with multiple classes or forces a large trim on all classes, which greatly reduces performance.

\section{Discussion} \label{s:discussion}

When run on the human-classified sources, the NN models had higher performance on the ACSVCS catalogue, whereas the RF models had higher performance on the NGVS testing catalogue. Of the NGVS test sources that were not classified correctly, approximately 95\% were contaminant sources classified as GCs+UCDs (false positives) while the other 5\% were GCs classified as contaminants (false negatives). This may suggest that either the level of confidence we expected for GCs classified by \cite{Oldham2016} was too high or that many of the sources they classify as interlopers are in fact GCs or UCDs.

\begin{figure}
    \centering
    \includegraphics[width=3.35in]{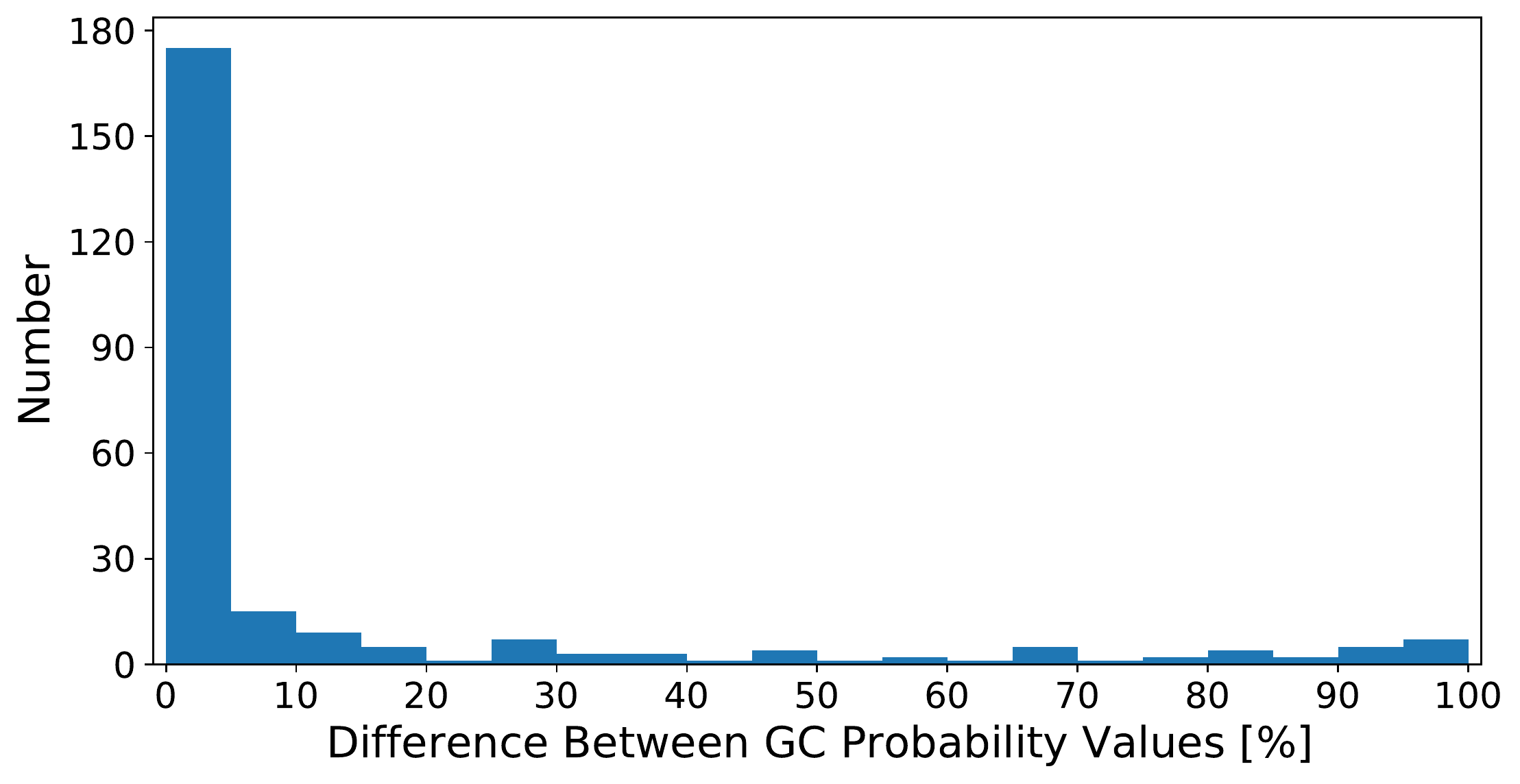}
    \caption{Absolute difference between pGC values from the ACSVCS catalogue and pRed + pBlue (pC) values from the NGVS test catalogue of sources found in both datasets. The classification of the majority of sources is agreed upon within 5\%, but there are also sources about which the two catalogues completely disagree (i.e., a 100\% difference in probability values).}
    \label{f:acsold}
\end{figure}

Both model architectures, however, perform better on the ACSVCS dataset which was taken by HST than the NGVS test dataset which comes from the same NGVS dataset as our training data (despite there being no overlap of sources). This is likely due to the robustness of selection criteria of the two human-selected datasets. \cite{Jordan2009} employ a more rigorous approach involving model-based clustering methods and catalogues of expected contaminants (as opposed to the colour-cuts used by \cite{Oldham2016}), which results in a more reliable classified list of sources. The higher spatial resolution of HST data also contributes to the reliability of human-classification. When comparing the two human-selected datasets, there are 253 sources in common and, while the majority of their GC classifications agree within 5\% of each other, there is some disagreement which may also increase the discrepancy in our models' performance (Fig.~\ref{f:acsold}).

\subsection{Limitations} \label{s:limits}

The main limitation in training our models is the limited number of confirmed sources, especially of UCD and background galaxy sources, so the models we created would benefit from further follow up analysis. GC populations have similar characteristics galaxy to galaxy (e.g., colour, size, and distribution across GC mass and luminosity), but in terms of apparent magnitude these features only appear similar for galaxies of similar distances. Therefore, while our use of absolute magnitudes allows our models to be run easily on other galaxies, it is necessary to limit their use to galaxies at similar distances to that of our training sample. Using our models on other well studied and well documented galaxies (such as the other giant ellipticals in Virgo, M49 and M60, which are well documented and at similar distances to M87) would provide an idea of how well they perform on sources outside of M87. It would be especially helpful to run NN3 and RF3 on larger samples since, as evidenced by their much poorer performance compared to the other models, we did not have access to enough UCD sources to create a successful UCD vs GC model. This is due to UCDs being particularly difficult to source and confirm (Section~\ref{s:ucdgc}). Creating more robust versions of these two models would also greatly aid further research into UCDs by efficiently creating more reliable candidate lists.

\begin{figure*}
    \centering
    \includegraphics[width=3.3in]{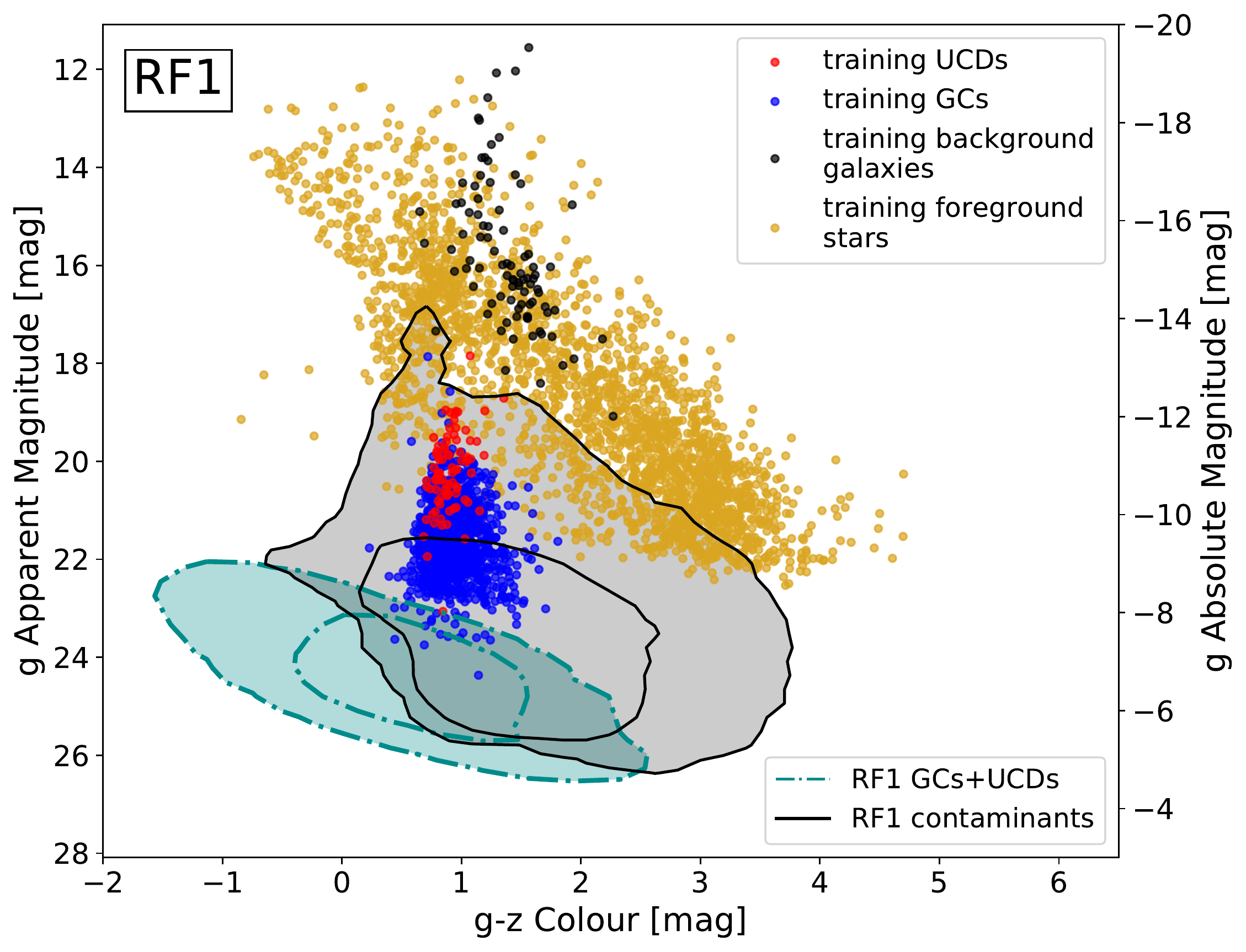}
    \includegraphics[width=3.3in]{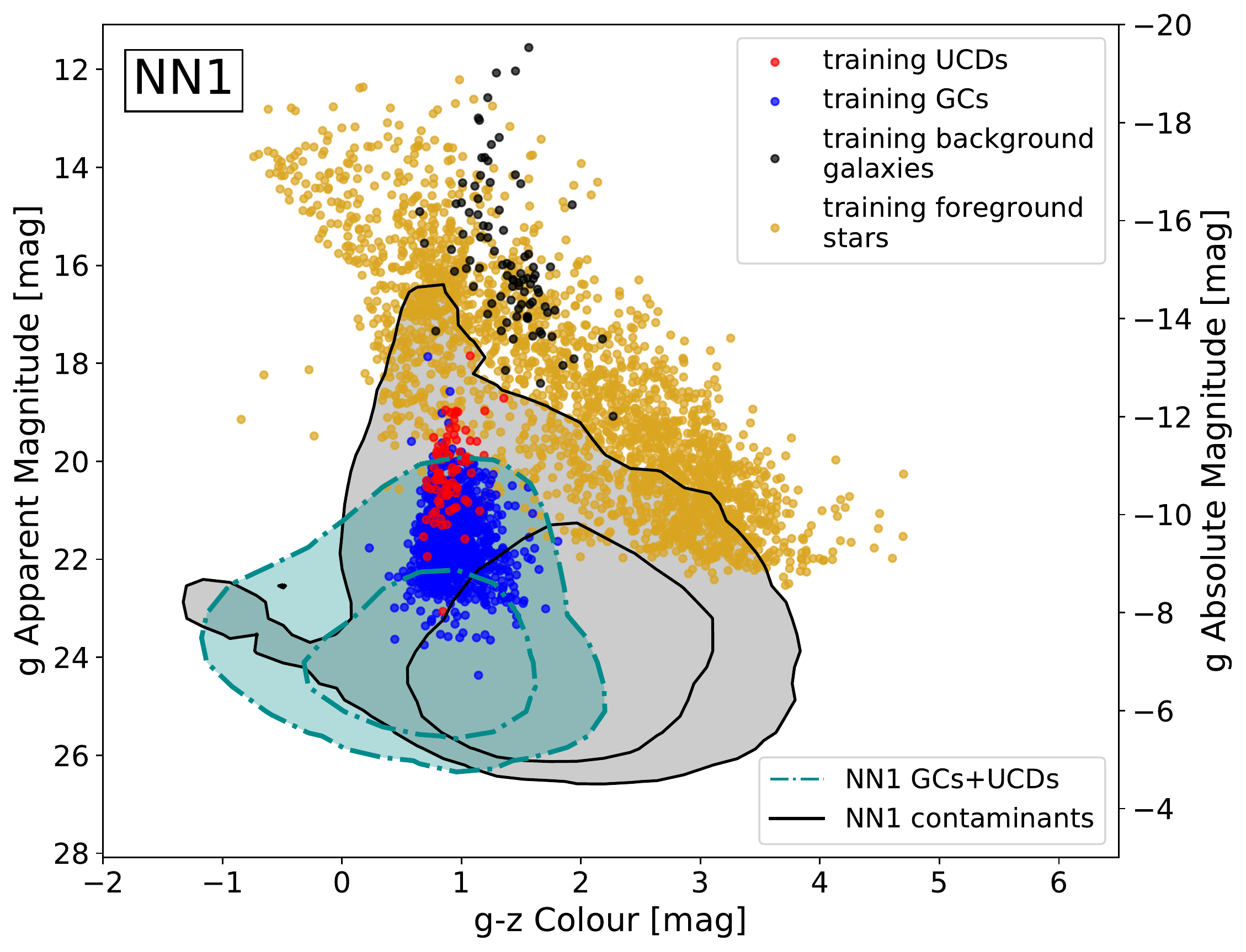}
    \includegraphics[width=3.3in]{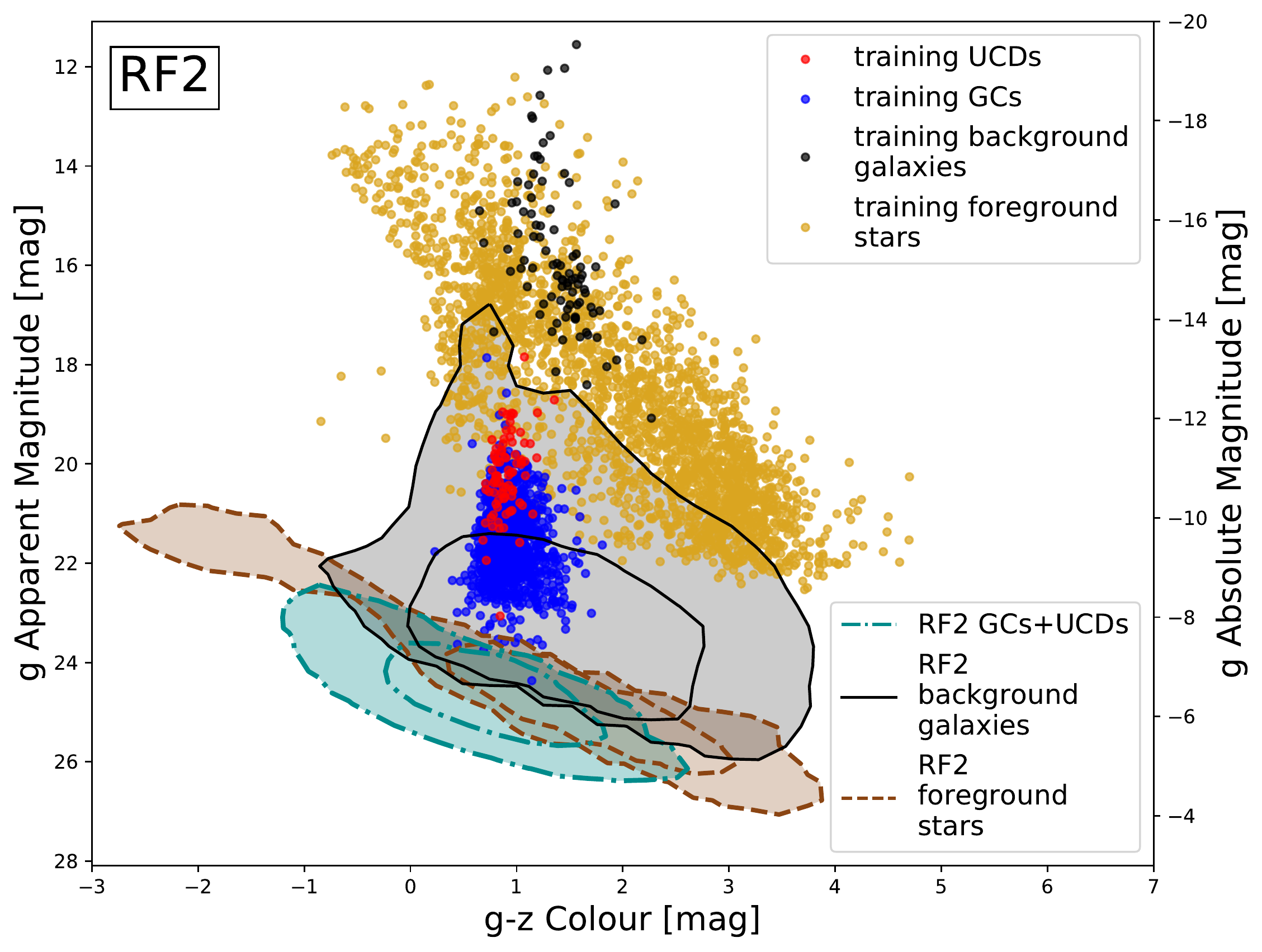}
    \includegraphics[width=3.3in]{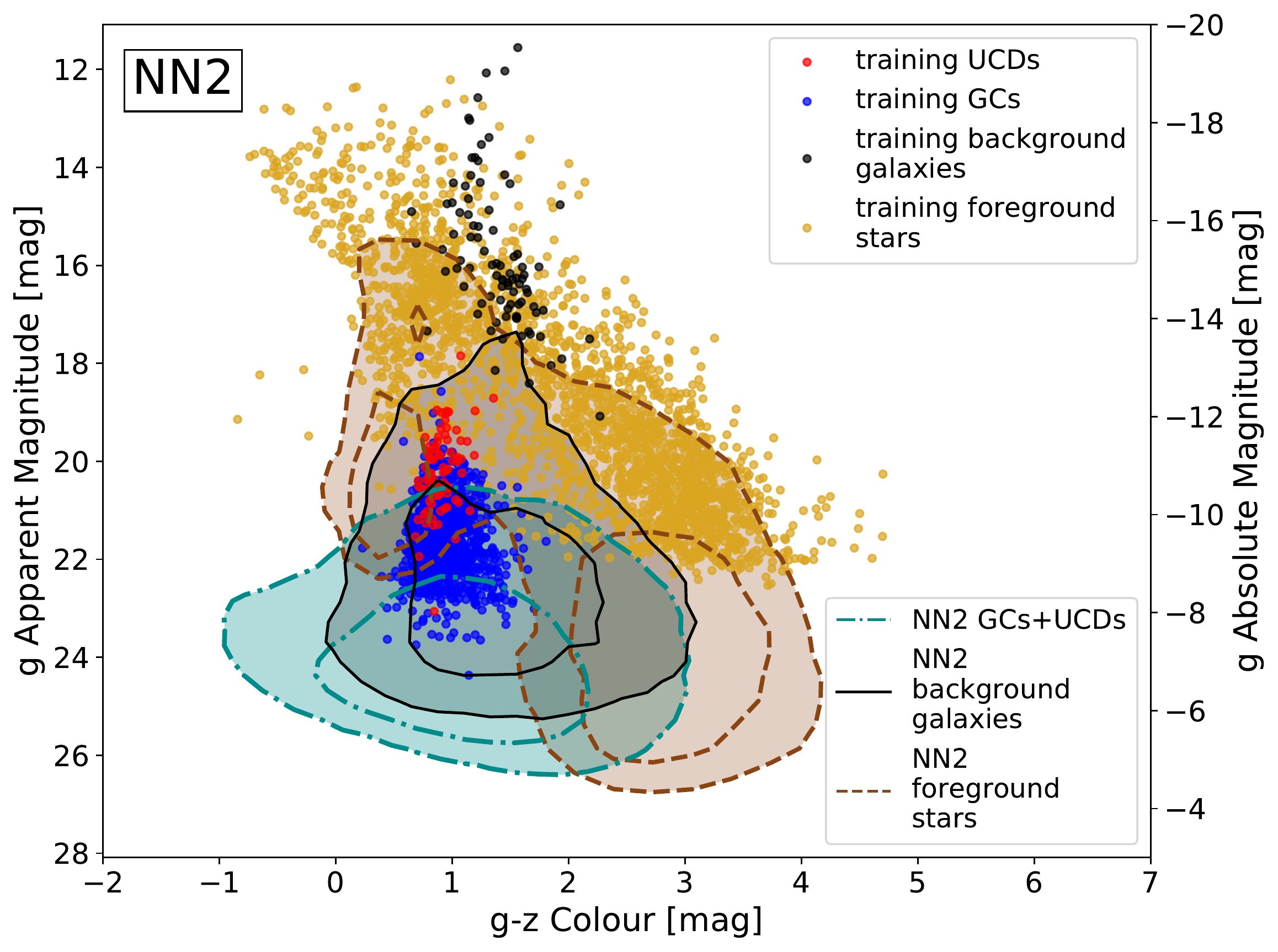}
    \includegraphics[width=3.3in]{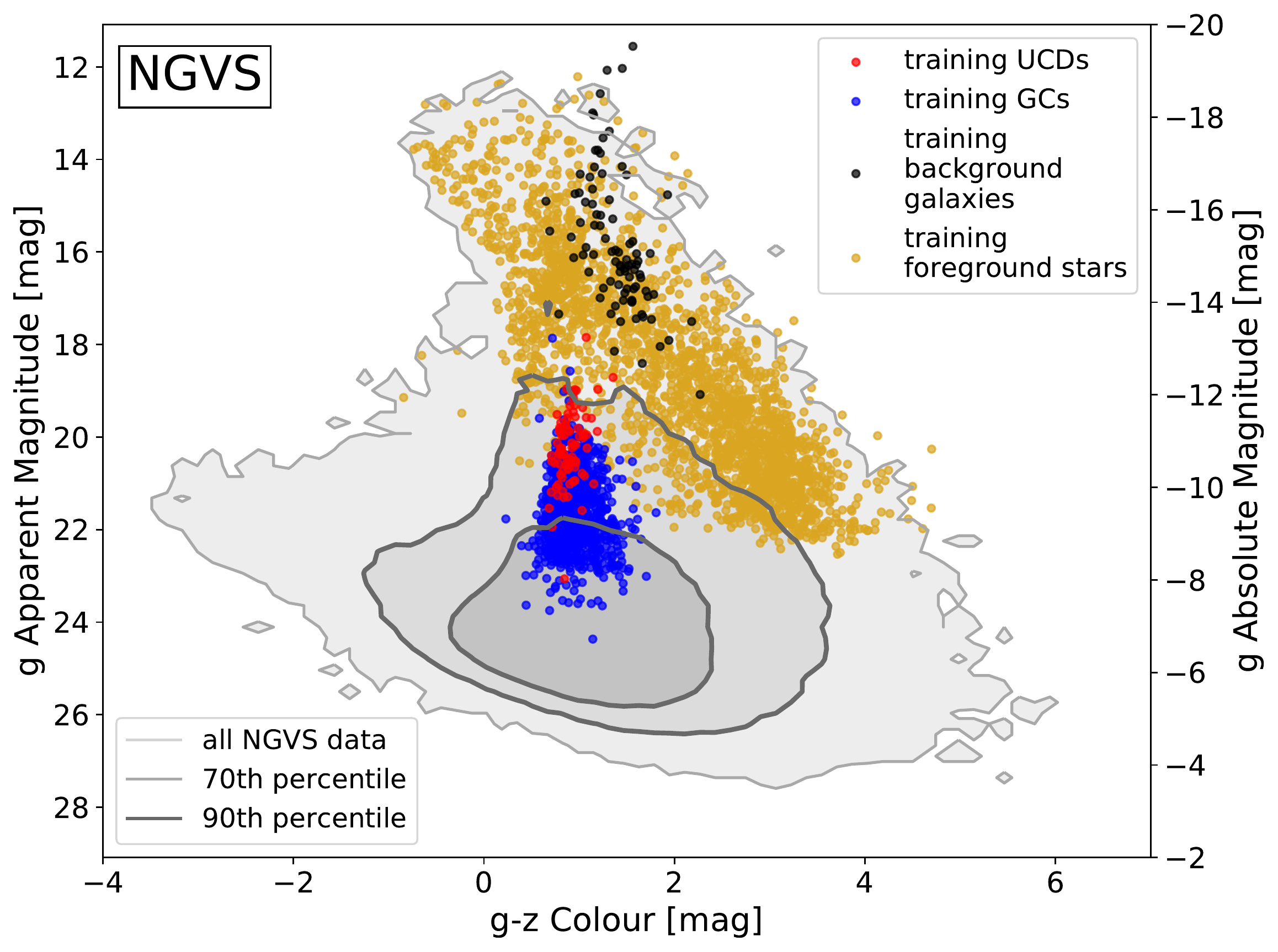}
    \caption{Colour-magnitude space of training data (scattered dots) compared to that of classified data (shaded contours) resulting from the RF models (upper left: RF1; lower left: RF2) and NN models (upper right: NN1; lower right: NN2) being run on the full, unfiltered NGVS catalogue of 719,600 sources (bottom centre). The contours show the 70th and 90th percentiles of non-zero values. The training data only consists of sources for which their classifications were spectroscopically confirmed and, as spectroscopy is typically only run on the brightest sources, the training data is much brighter than the rest of the data in the full NGVS dataset. Evidently, the models do still classify fainter sources and, despite some overlap, the different populations that each model extracts are distinct.}
    \label{f:colmagNGVS}
\end{figure*}

A major concern when creating these models was whether they would be able to accurately sort through sources with fainter magnitudes, especially considering that \cite{Thilker2022} were able to classify sources $\sim$1 mag fainter by using ML than by using human classification alone. Our models were only trained on sources for which their classifications were spectroscopically confirmed and, as spectroscopy is typically only run on the brightest sources, our training data is much brighter than the rest of the data in the full NGVS dataset. For this reason we ran our four main models (RF1, RF2, NN1, and NN2) on the full, unfiltered dataset of 719,600 NGVS sources to see how our models classified fainter sources. When viewed in colour-magnitude space (Fig.~\ref{f:colmagNGVS}), it’s clear that they do still classify fainter sources and, despite some overlap, the different populations that each model extracts are distinct. The four models classify each population differently (especially RF2 and NN2), but given that they all had high performance --- despite the RF models performing much better on NGVS test data and the NN models performing much better on the ACSVCS data --- further investigation and larger tests are required before concluding which is the better model algorithm.

\subsection{Advantages} \label{s:advantages}

Our approach of using solely photometric data from a ground-based telescope is important for two main reasons. The first being that incorporating space-based HST data would involve using non-uniform observations since the telescope uses a mix of different filters and exposure lengths for different galaxies. Since its camera is often pointed directly at galactic centres, this means we would be neglecting GCs in the outer halo in our training data. The second reason is that Rubin Observatory's upcoming LSST survey will offer a uniform dataset and our results show it is possible to train a model using only photometric data in the same formats and filters as will be available. LSST is also expected to be a large survey, with a resulting very large amount of data and, hence, our goal is to create tools that can help sift through all these new observations efficiently. Our models are, however, intended to be only that: tools that generate robust, yet unconfirmed candidate lists that still require targeted follow-up inspection and analysis, much like \cite{Saifollahi2021} did with their models generating UCD candidates. The two model types chosen, RF and NN, are two of the more common ML algorithms with minimal barrier to entry for ML beginners that are suited for our supervised classification task. This makes our models and methods more accessible to those who wish to perform follow-up analysis or start a similar project. Given the advantages of using ML on large surveys, this proves to be a promising method of GC and UCD candidate generation, provided the appropriate care is taken in selecting training sets and inputs features.

\subsection{Recommendations for Future Use} \label{s:future}

Aside from training models on larger datasets, another important consideration for follow-up research is that our models are not linearly independent, in that some features (in this case, the 4 colour features) are linear combinations of other features (i.e., magnitudes). Generally, ML algorithms may perform differently depending on whether or not features are linearly independent, as this can sometimes emphasize the importance of dependant features or make them redundant. When creating our models, we found by trial and error that most of our models (with the exception of NN3) performed better with both colour and magnitude features, but with future improvement of these models it will be important to verify or reevaluate this choice. Unlike NN packages, the \texttt{RandomForestClassifier} class has a built-in attribute, \texttt{feature\_importances\_}, which indicates impurity-based feature importances, also known as the Gini importance \citep{scikit-learn}. During follow-up analysis we accessed this attribute wherein which the importance of each feature is calculated based on how much it reduces the criterion (i.e., the function that measures the quality of each new branch in RF trees).  We found that the magnitude and flux radius features were most important, while the colour features were comparatively less important. Note that this does not imply that colour features are not important, simply that they may not be as important as magnitude features, since colours show the relationships between magnitude bands and models will assess features relationships during training. This, therefore, may suggest that while the colour features do seem to improve model performance, they are likely not essential for a successful random forest model due to linear dependence on magnitude features. This also suggests that the flux radius features do help the models distinguish background sources from other sources, as these are expected to be less spatially resolved.


\section{Conclusion} \label{s:conclusion}

With the large volume of data from upcoming sky surveys, we aimed to create ML classification tools that will create lists of GC+UCD candidates by running on datasets of photometric measurements alone. We employ two supervised ML algorithms: random forest and neural networks, and three architectures, two of which focus on selecting GCs+UCDs from unfiltered catalogues and one auxiliary one which focuses on selecting UCDs from catologues of only star cluster sources. The two main architectures have promising performance, but the auxiliary models will likely require more confirmed UCD sources before reaching optimal performance levels. When compared to the ACSVCS dataset of human-selected GCs, the best performing random forest model is able to reselect 61.2\% $\pm$ 8.0\% of GCs and the best performing neural network model reselects 95.0\% $\pm$ 3.4\% of them. When compared to GCs and interlopers of the NGVS test set, the random forest models can correctly classify 91.0\% $\pm$ 1.2\% of them and the neural network models can correctly classify 57.3\% $\pm$ 1.1\%. Note that there are inherently more systematic uncertainties in the human-classification method, which may contribute to poorer performance when testing the NN models. Additionally, the strength of our NN method is indicated by achieving 62\% as good classification on seeing-limited data compared to HST data.

We show in this paper that existing and accessible ML techniques can be used to successfully classify objects in large-scale photometric surveys. Others have attempted this problem with galaxies at different distances with different algorithms and features and have achieved similar success, which supports the robustness of this technique \citep{Mohammadi2022}. Anticipating the first light of Rubin Observatory, we have created tools that will properly run on upcoming data products. Future development of ML algorithms like ours only serve to improve the accuracy and easy with which it will be possible to search for important and scientifically illuminating sources, such as GCs and UCDs.


\section*{Acknowledgements}
EB, KCD, and DH acknowledge funding from the Natural Sciences and Engineering Research Council of Canada (NSERC), the Canada Research Chairs (CRC) program, and the McGill Bob Wares Science Innovation Prospectors Fund. KCD acknowledges fellowship funding from the McGill Space Institute. AK acknowledges support from NASA through grant number GO-14738 from STSci. We thank Jay Strader for sharing a preliminary version of the M87 spectroscopic catalog. We also thank the referee for their helpful comments which greatly improved this work.

This publication makes use of data products from the Two Micron All Sky Survey, which is a joint project of the University of Massachusetts and the Infrared Processing and Analysis Center/California Institute of Technology, funded by the National Aeronautics and Space Administration and the National Science Foundation.

This publication is based on observations obtained with MegaPrime/MegaCam, a joint project of CFHT and CEA/DAPNIA, at the Canada-France-Hawaii Telescope (CFHT) which is operated by the National Research Council (NRC) of Canada, the Institut National des Sciences de l'Univers of the Centre National de la Recherche Scientifique of France, and the University of Hawaii.

The Legacy Surveys consist of three individual and complementary projects: the Dark Energy Camera Legacy Survey (DECaLS; Proposal ID \#2014B-0404; PIs: David Schlegel and Arjun Dey), the Beijing-Arizona Sky Survey (BASS; NOAO Prop. ID \#2015A-0801; PIs: Zhou Xu and Xiaohui Fan), and the Mayall z-band Legacy Survey (MzLS; Prop. ID \#2016A-0453; PI: Arjun Dey). DECaLS, BASS and MzLS together include data obtained, respectively, at the Blanco telescope, Cerro Tololo Inter-American Observatory, NSF’s NOIRLab; the Bok telescope, Steward Observatory, University of Arizona; and the Mayall telescope, Kitt Peak National Observatory, NOIRLab. The Legacy Surveys project is honored to be permitted to conduct astronomical research on Iolkam Du’ag (Kitt Peak), a mountain with particular significance to the Tohono O’odham Nation.

NOIRLab is operated by the Association of Universities for Research in Astronomy (AURA) under a cooperative agreement with the National Science Foundation.

This project used data obtained with the Dark Energy Camera (DECam), which was constructed by the Dark Energy Survey (DES) collaboration. Funding for the DES Projects has been provided by many sources detailed here: \url{https://noirlab.edu/science/about/scientific-acknowledgments#decals}.

BASS is a key project of the Telescope Access Program (TAP), which has been funded by the National Astronomical Observatories of China, the Chinese Academy of Sciences (the Strategic Priority Research Program “The Emergence of Cosmological Structures” Grant \# XDB09000000), and the Special Fund for Astronomy from the Ministry of Finance. The BASS is also supported by the External Cooperation Program of Chinese Academy of Sciences (Grant \# 114A11KYSB20160057), and Chinese National Natural Science Foundation (Grant \# 11433005).

The Legacy Survey team makes use of data products from the Near-Earth Object Wide-field Infrared Survey Explorer (NEOWISE), which is a project of the Jet Propulsion Laboratory/California Institute of Technology. NEOWISE is funded by the National Aeronautics and Space Administration.

The Legacy Surveys imaging of the DESI footprint is supported by the Director, Office of Science, Office of High Energy Physics of the U.S. Department of Energy under Contract No. DE-AC02-05CH1123, by the National Energy Research Scientific Computing Center, a DOE Office of Science User Facility under the same contract; and by the U.S. National Science Foundation, Division of Astronomical Sciences under Contract No. AST-0950945 to NOAO.

This work has made use of data from the European Space Agency (ESA) mission {\it Gaia} (\url{https://www.cosmos.esa.int/gaia}), processed by the {\it Gaia} Data Processing and Analysis Consortium (DPAC, \url{https://www.cosmos.esa.int/web/gaia/dpac/consortium}). Funding for the DPAC has been provided by national institutions, in particular the institutions participating in the {\it Gaia} Multilateral Agreement.

\section*{Data Availability}
The CFHT data used in this study is publicly available through the CADC (\url{https://www.cadc-ccda.hia-iha.nrc-cnrc.gc.ca/en/}). We also source data from the following catalogues: \cite{Jordan2009}, \cite{Oldham2016}, \cite{Pandya2016}, \cite{Zhang2015}, and Strader et al. (in prep.); and from the following catalogues queried through the databases on NOIRLab’s Astro Data Lab (\url{https://datalab.noirlab.edu}): \cite{Skrutskie2006}, \cite{Dey2019}, and \cite{Gaia2018}.



\bibliographystyle{mnras}
\bibliography{refs}


\newpage
\appendix

\section{CFHT Observations of M87} \label{s:appendixA}
M87 is in the field of many MegaPipe observations. These are itemised in Table \ref{t:ngvsobs}.

\begin{table}
    \centering
    \caption{CADC MegaPipe observations covering M87.}
    \begin{tabular}{ |c|c|c| }
        \hline
        MegaPipe.364.203 & \makecell{MegaPipe.365.201 \\ MegaPipe.365.202 \\ MegaPipe.365.203 \\ MegaPipe.365.206} & \makecell{MegaPipe.366.201 \\ MegaPipe.366.202 \\ MegaPipe.366.203 \\ MegaPipe.366.204 \\ MegaPipe.366.205 \\ MegaPipe.366.206}\\
        \hline
        \makecell{MegaPipe.367.201 \\ MegaPipe.367.202 \\ MegaPipe.367.203 \\ MegaPipe.367.204 \\ MegaPipe.367.205 \\ MegaPipe.367.206} & \makecell{MegaPipe.368.200 \\ MegaPipe.368.201 \\ MegaPipe.368.202 \\ MegaPipe.368.203 \\ MegaPipe.368.204 \\ MegaPipe.368.205 \\ MegaPipe.368.206} & \makecell{MegaPipe.369.198 \\ MegaPipe.369.199 \\ MegaPipe.369.200 \\ MegaPipe.369.202 \\ MegaPipe.369.203 \\ MegaPipe.369.204 \\ MegaPipe.369.205 \\ MegaPipe.369.206}\\
        \hline
        \makecell{MegaPipe.370.198 \\ MegaPipe.370.199 \\ MegaPipe.370.200 \\ MegaPipe.370.202 \\ MegaPipe.370.203 \\ MegaPipe.370.204 \\ MegaPipe.370.205 \\ MegaPipe.370.206} & \makecell{MegaPipe.371.198 \\ MegaPipe.371.199 \\ MegaPipe.371.200 \\ MegaPipe.371.204} & MegaPipe.372.198\\
        \hline
    \end{tabular}
    \label{t:ngvsobs}
\end{table}

\section{Confusion matrices} \label{s:appendixB}

\begin{figure*}
    \centering
    \includegraphics[width=3.3in]{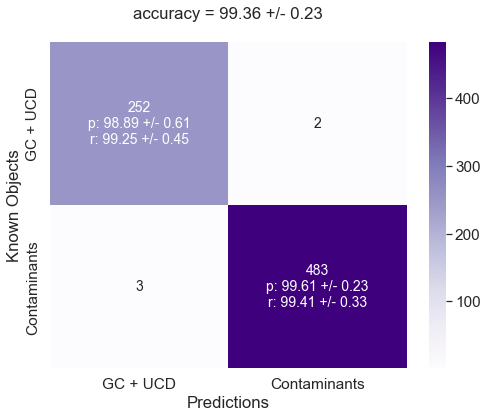}
    \includegraphics[width=3.3in]{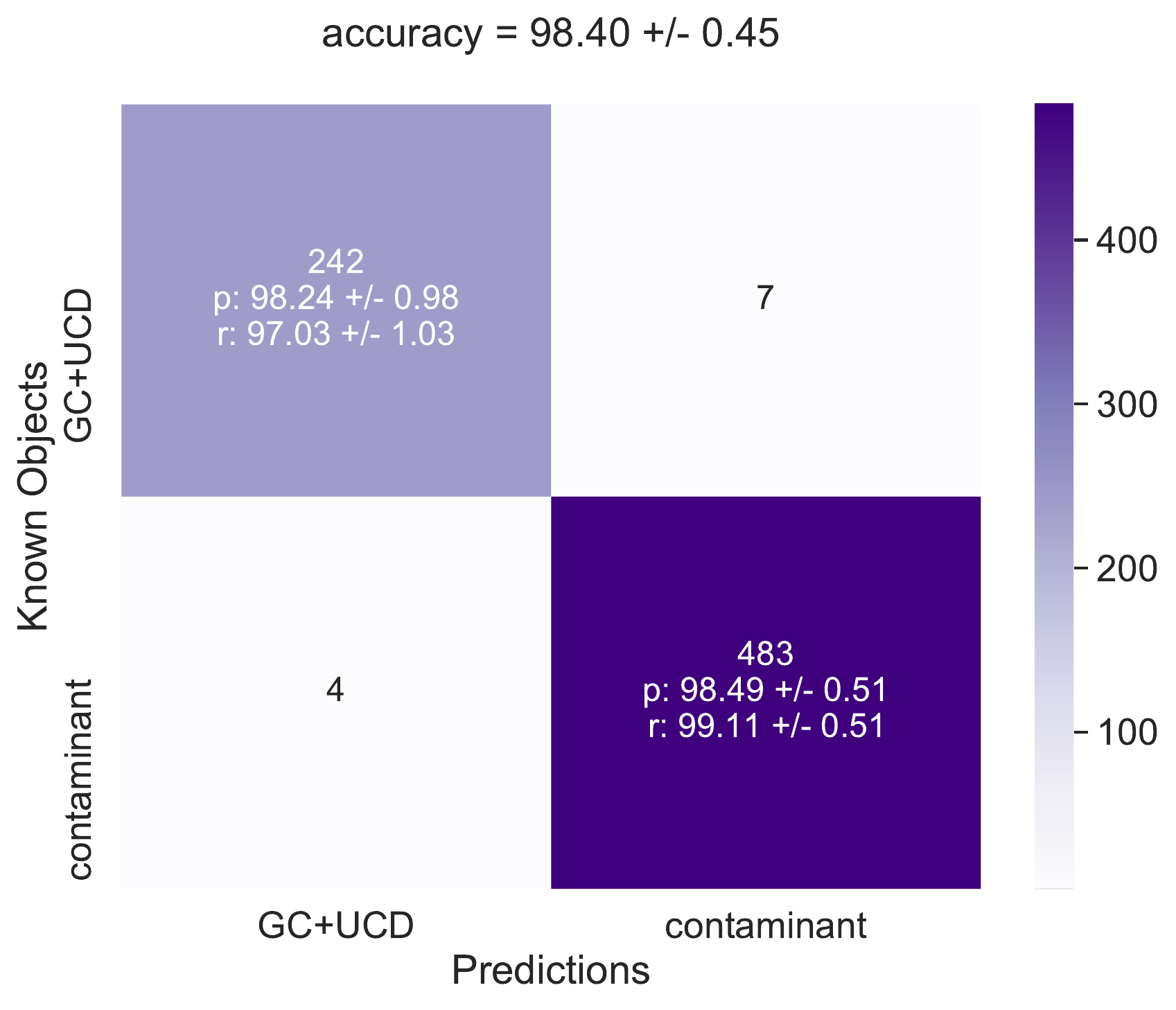}
    \includegraphics[width=3.3in]{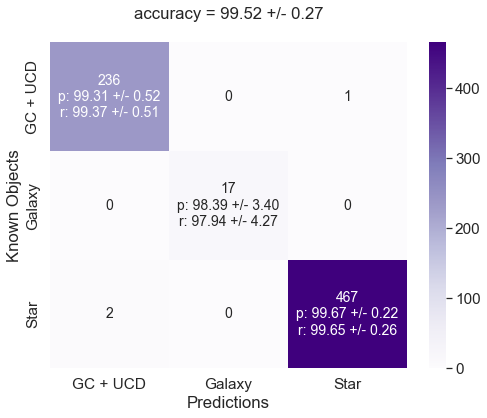}
    \includegraphics[width=3.3in]{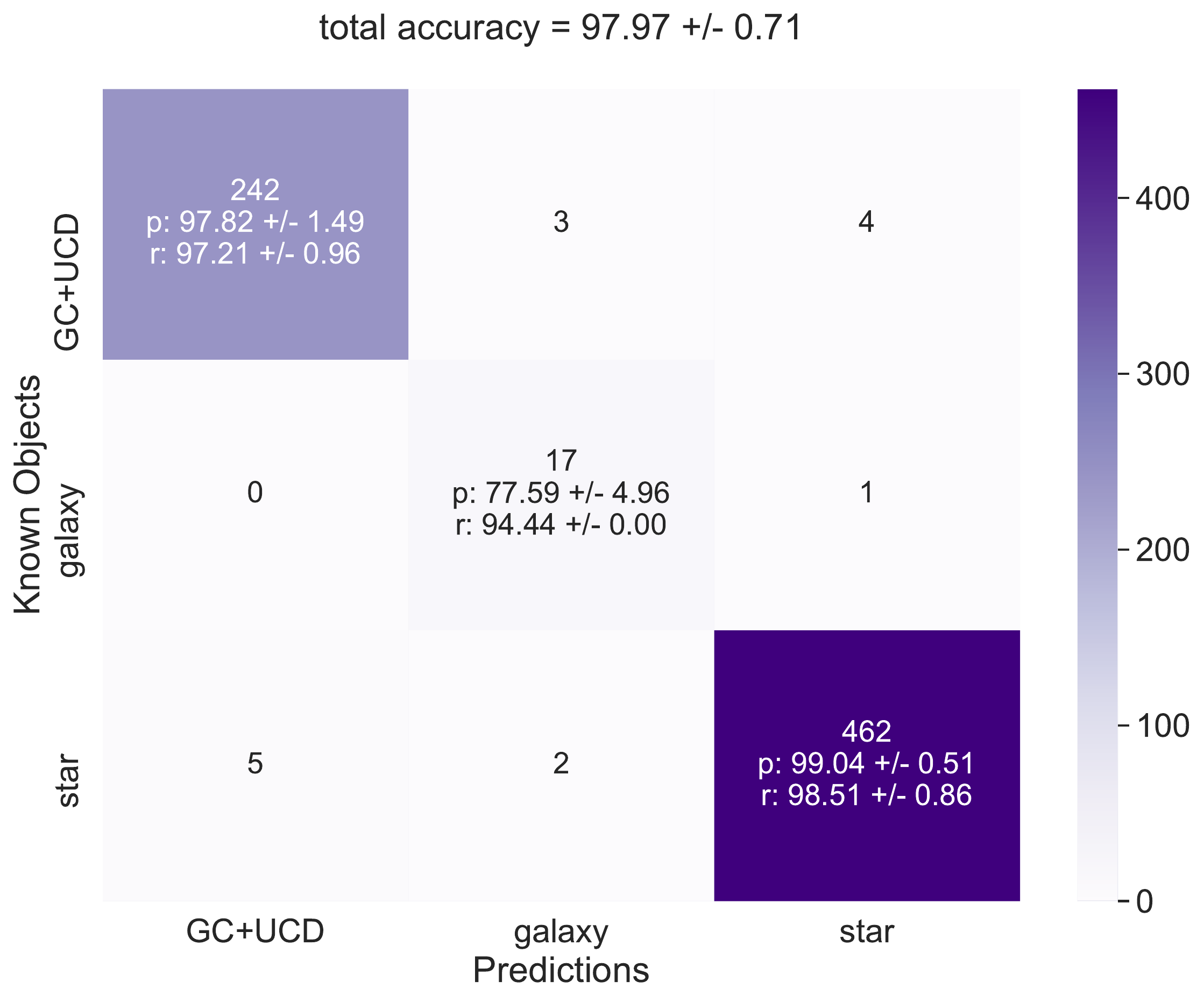}
    \includegraphics[width=3.3in]{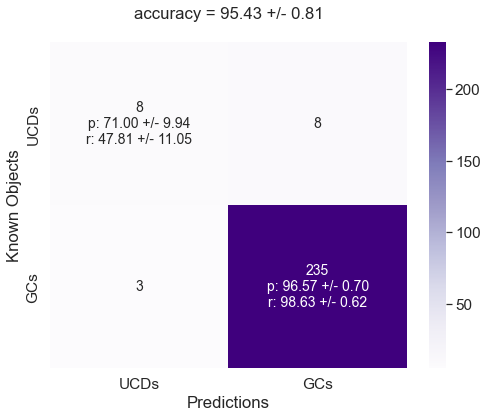}
    \includegraphics[width=3.7in]{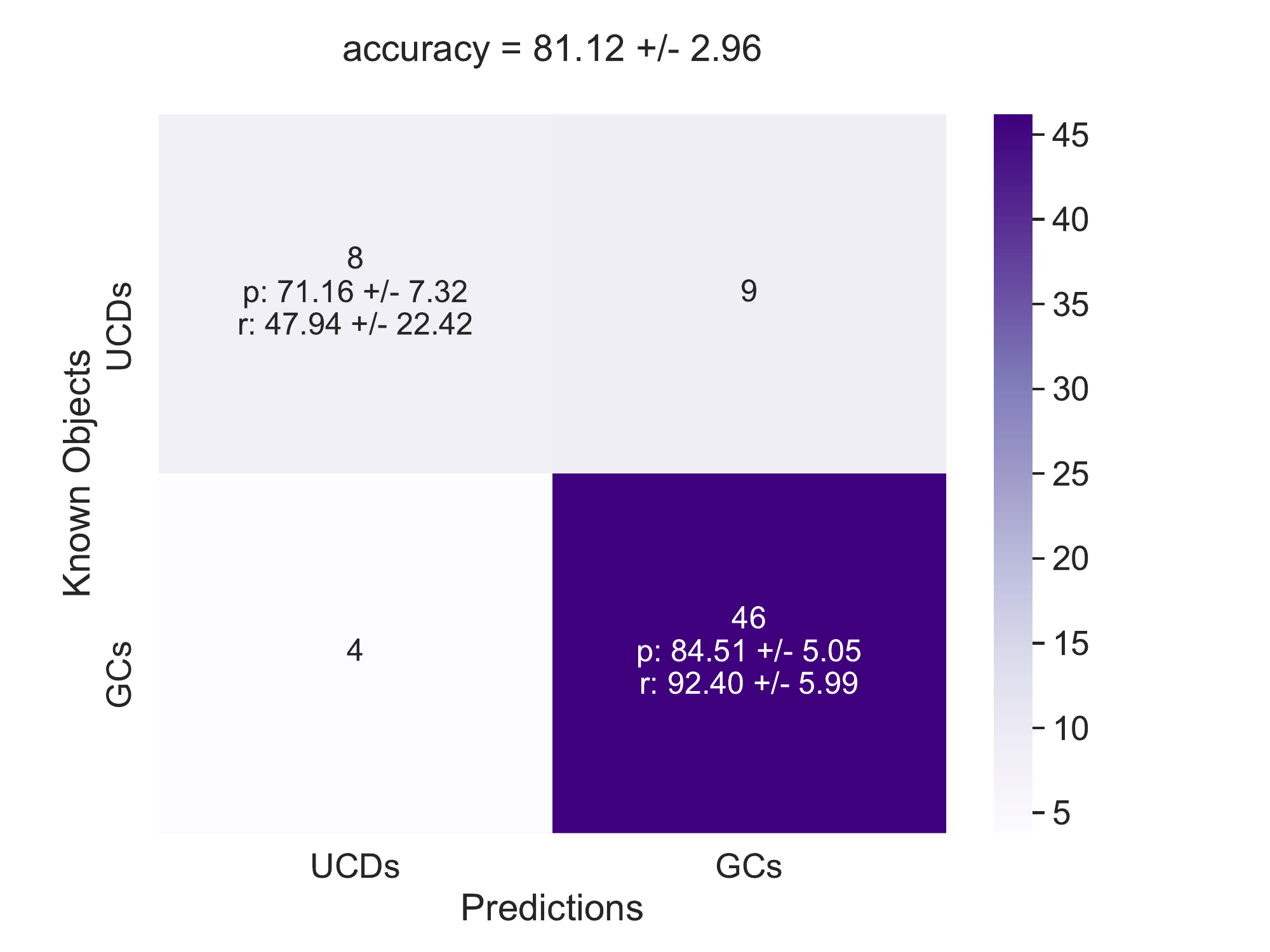}
    \caption{Confusion matrices for all six models developed in this work, positioned as follows: top-left: RF1; top-right: NN1; middle-left: RF2; middle-right: NN2; lower-left: RF3; lower-right: NN3. Overall accuracy is listed at the top of each figure and counts of classified sources are listed on the matrices, with precision and recall values for each class denoted as "p" and "r" respectively.}
    \label{f:cm}
\end{figure*}


\bsp	
\label{lastpage}
\end{document}